\documentclass[prd,amsmath,amssymb,showpacs,showkeys,floatfix,
               nofootinbib,eqsecnum,
               preprint,12pt,tightenlines,fleqn
               ]{revtex4-1}
\usepackage{graphicx}

\newcommand{\beq}{\begin{equation}}
\newcommand{\eeq}{\end{equation}}
\newcommand{\bea}{\begin{eqnarray}}
\newcommand{\eea}{\end{eqnarray}}
\newcommand{\bsubeqs}{\begin{subequations}}
\newcommand{\esubeqs}{\end{subequations}}

\newcommand{\YMH}  {Yang--Mills--Higgs}
\newcommand{\YMHth}{Yang--Mills--Higgs theory}

\newcommand{\no}{\nonumber}
\newcommand{\bR}{\mathbb{R}}

\newcommand{\frachalf}{{\textstyle \frac{1}{2}}}
\newcommand{\dvx}{\!{\rm d}^3 x\,\,}

\newcommand{\dint}[1]{\!{\rm d}#1\,\,}

\newcommand{\tr}{\text{tr}}

\newcommand{\vect}     [1]{\left( \begin{array}{c} #1 \end{array} \right)}
\newcommand{\vectleft} [1]{\left( \begin{array}{l} #1 \end{array} \right)}

\newcommand{\threematw}[1]{\left( \begin{array}{c@{\hspace{4mm}}c@{\hspace{4mm}}c} #1 \end{array} \right)}

\begin{document}

\noindent  Phys. Rev. D  96, 016006 (2017) 
\hfill arXiv:1704.07756  %%\;(\version)
\newline\vspace*{2mm}
%
%\noindent  arXiv:1704.07756
%\hfill  KA--TP--18--2017\;(\version)
%\newline\vspace*{2mm}
%

\title{$SU(3)$ sphaleron: Numerical solution}
\author{F.R. Klinkhamer}
\email{frans.klinkhamer@kit.edu}
\author{P. Nagel}
\email{pascal.nagel@kit.edu}
\affiliation{Institute for
Theoretical Physics, Karlsruhe Institute of
Technology (KIT), 76128 Karlsruhe, Germany\\}
\begin{abstract}
\noindent
\vspace*{0mm}\newline
We complete the construction of the sphaleron $\widehat{S}$
in $SU(3)$ Yang-Mills-Higgs theory with a single Higgs triplet
by solving the reduced field equations numerically.
The energy of the $SU(3)$ sphaleron $\widehat{S}$
is found to be of the same order as the energy of a previously known
solution, the embedded $SU(2)\times U(1)$ sphaleron $S$.
In addition, we discuss $\widehat{S}$
in an extended $SU(3)$
Yang-Mills-Higgs theory with three Higgs triplets,
where all eight gauge bosons get an equal mass in the vacuum.
This extended $SU(3)$ Yang-Mills-Higgs theory may be considered
as a toy model of
quantum chromodynamics without quark fields
and we conjecture that the
$\widehat{S}$ gauge fields play a significant role in the nonperturbative
dynamics of quantum chromodynamics
(which does not have fundamental scalar fields but gets a mass scale
from quantum effects).
\end{abstract}

%\pacs{11.30.Cp, 12.20.-m}
%\keywords{Lorentz violation, quantum electrodynamics}

%PhySH Concepts:
%Classical solutions in field theory (Primary)|Glueballs

\maketitle

%%\newpage%%tmp
\section{Introduction}
\label{sec:Introduction}

The non-Abelian chiral gauge anomaly~\cite{Bardeen1969}
is expected to be associated~\cite{Klinkhamer1998}
with a new type of sphaleron
(a static, but unstable, finite-energy solution of the
classical field equations).
A self-consistent \textit{Ansatz} for this sphaleron,
denoted $\widehat{S}$, has indeed been
constructed in $SU(3)$ \YMH~theory~\cite{KlinkhamerRupp2005}.
But the numerical solution of the reduced field equations and
the corresponding determination of the energy $E_{\widehat{S}}$
have turned out to be challenging.
In this article, we present, at last, the numerical
solution of the $\widehat{S}$ fields in the basic $SU(3)$ \YMH~theory
with a single Higgs triplet and find a surprisingly low
value of the energy $E_{\widehat{S}}$,
namely an energy of the same order as
(and even below) the energy $E_{S}$ of the embedded
$SU(2)\times U(1)$ sphaleron
$S$~\cite{KlinkhamerManton1984,KlinkhamerLaterveer1990,%
KunzKleihausBrihaye1992}.

The outline of the present article is as follows.
In Sec.~\ref{sec:Two-SU3-YMH-Theories},
we define two classical $SU(3)$ \YMH~theories. The first theory
has a single Higgs triplet and the second theory
has three Higgs triplets
(designed to give an equal mass to all eight gauge bosons in the vacuum).
The focus of the main part of this article will be on
the basic $SU(3)$ \YMH~theory with a single Higgs triplet.
In Sec.~\ref{sec:Shat-Ansatz}, we give a brief sketch of the
topological argument (minimax procedure) and recall the
$\widehat{S}$ \textit{Ansatz} from Ref.~\cite{KlinkhamerRupp2005}.
In Sec.~\ref{sec:Field-equations-analytical-results},
we consider the reduced field equations and solve
them analytically near the origin.
In Sec.~\ref{sec:Numerical-results},
we present the numerical solution obtained
by a minimization procedure of the \textit{Ansatz} energy.
In Sec.~\ref{sec:Shat-in-extended-SU3-YMHth},
we give the corresponding results for
$\widehat{S}$  in the extended $SU(3)$ \YMH~theory
with three Higgs triplets.
In Sec.~\ref{sec:Conclusion}, we present concluding remarks.

There are also five appendices with technical details.
For the basic $SU(3)$ \YMH~theory,
Appendix~\ref{app:Shat-energy-density-basic-YMHth}
gives the $\widehat{S}$ energy density
and Appendix~\ref{app:Expansion-coefficients-basic-YMHth}
presents the expansion coefficients
for the $\widehat{S}$  \textit{Ansatz} functions.
For the extended $SU(3)$ \YMH~theory,
Appendix~\ref{app:NCS-in-extended-SU3-YMHth} presents the
noncontractible sphere of configurations
needed for the $\widehat{S}$ \textit{Ansatz},
Appendix~\ref{app:Shat-energy-density-extended-SU3-YMHth}
gives the $\widehat{S}$ energy density,
and Appendix~\ref{app:Minimization-setup-extended-SU3-YMHth}
discusses the minimization setup.

%%\newpage%%tmp
\section{Two $SU(3)$ Yang--Mills--Higgs theories}
%%from "More on the $SU(3)$ sphaleron.." (March 4, 2017  v0.46frk)
\label{sec:Two-SU3-YMH-Theories}

We consider two classical $SU(3)$ Yang--Mills--Higgs (YMH) theories.
The first theory is a direct enlargement~\cite{Weinberg1972}
of the $SU(2)\times U(1)$  electroweak Standard Model
with weak mixing angle $\theta_{w}=\pi/6$.
The second theory may be considered as a toy model of
a simplified version of
quantum chromodynamics (QCD)~\cite{PDG2016}
without quark fields,
having eight gauge bosons of equal mass
(taken to model the quantum effects of QCD).
Some further remarks on the possible relevance of the
the second $SU(3)$ YMH theory for quarkless QCD are presented in
Sec.~\ref{subsec:Discussion-in-extended-SU3-YMHth}.
The first $SU(3)$ \YMHth~is the one used in the
original $\widehat{S}$ paper~\cite{KlinkhamerRupp2005}
and will be the main focus of the present article.

\subsection{Basic $SU(3)$ YMH theory}
\label{subsec:Embedded-electroweak-theory}

The first $SU(3)$ \YMHth~considered has a single triplet $\Phi$
of complex scalar fields. The classical action is given by
\begin{equation}
S = \int_{\bR^4}\; d^4x
\left\{ \frachalf \, \tr\, F_{\mu\nu}  F^{\mu\nu} +
       (D_\mu  \Phi)^\dagger \,(D^\mu \Phi)     -
       \lambda \, \left( \Phi^\dagger \Phi - \eta^2 \right)^2
 \right\}\,,
\label{eq:actionYMH1}
\end{equation}
where $F_{\mu\nu}\equiv\partial_\mu A_\nu -\partial_\nu A_\mu + g [A_\mu,
A_\nu]$ is the  $SU(3)$  Yang--Mills field strength tensor
and $D_\mu \equiv (\partial_\mu + g A_\mu)$
the covariant derivative for the triplet representation
of $SU(3)$. The Higgs field has a global $U(1)$ symmetry,
$\Phi(x)\to e^{i\omega}\, \Phi(x)$.
The constant $\eta$ is assumed to be nonzero
and the standard electroweak notation is obtained by setting
$\eta=v/\sqrt{2}$.

The $SU(3)$ Yang--Mills gauge field is defined as
\begin{equation}
A_\mu (x) \equiv A_\mu^a (x)\, \lambda_a / (2 i )\,,
\label{Adef}
\end{equation}
in terms of the eight Gell-Mann matrices
\begin{align}
\lambda_1 &=
\threematw{0 & 1 & 0 \\[-1ex] 1 & 0 & 0 \\[-1ex] 0 & 0 & 0 }\,, &
\lambda_2 &=
\threematw{0 & - i  & 0 \\[-1ex]  i  & 0 & 0 \\[-1ex] 0 & 0 & 0 }\,, &
\lambda_3 &=
\threematw{1 & 0 & 0 \\[-1ex] 0 & -1 & 0 \\[-1ex] 0 & 0 & 0 }\,,
\nonumber\\[1em]
\lambda_4 &=
\threematw{0 & 0 & 1 \\[-1ex] 0 & 0 & 0 \\[-1ex] 1 & 0 & 0 }\,, &
\lambda_5 &=
\threematw{0 & 0 & - i  \\[-1ex] 0 & 0 & 0 \\[-1ex]  i  & 0 & 0 }\,,&
\lambda_6 &=
\threematw{0 & 0 & 0 \\[-1ex] 0 & 0 & 1 \\[-1ex] 0 & 1 & 0 }\,, &
\nonumber \\[1em]
\lambda_7 &=
\threematw{ 0 & 0 & 0 \\[-1ex] 0 & 0 & - i  \\[-1ex] 0 &  i  & 0}\,,&
\lambda_8 &=  \frac{1}{\sqrt{3}}
\threematw{1 & 0 & 0 \\[-1ex] 0 & 1 & 0 \\[-1ex] 0  & 0 & -2 }\,.
\end{align}
The field $\Phi(x)$ is a triplet of complex scalar fields,
\begin{equation}
\Phi(x)= \left(
          \begin{array}{c}
             \Phi_1(x)\\
             \Phi_2(x)\\
             \Phi_3(x)\\
          \end{array}
        \right)\,,
\end{equation}
which acquires a vacuum expectation value $\eta$
due to the Higgs potential term in the action \eqref{eq:actionYMH1}.
Throughout, we use the Minkowski spacetime metric
$g_{\mu\nu}(x)=\eta_{\mu\nu}=
[\textrm{diag}(+1,\,-1,\,-1,\,-1)]_{\mu\nu}$
and natural units with $\hbar=c=1$.

The scalar vacuum field can be chosen as
\begin{equation}
\Phi= \left(
          \begin{array}{c}
            0 \\
            0 \\
            \eta \\
          \end{array}
        \right)\,,
\end{equation}
which gives a mass to five gauge fields,
$ A_\mu^a$  for $a=4,\,5,\,6,\,7,\,8$,
with three gauge fields remaining massless,
$ A_\mu^a$ for $a=1,\,2,\,3$.
There is one physical scalar mode ($3\times 2 - 5 =1$), which is massive
for a nonvanishing quartic Higgs coupling, $\lambda >0$.
Equivalent Higgs vacua can, for example, be obtained by transformation
with the following $SU(3)$ matrices:
\begin{equation}
\label{eq:M1-M2-M3}
M_1 \equiv
\left(
  \begin{array}{ccc}
    1 & 0  & 0 \\
    0 & 0  & 1 \\
    0 & -1 & 0 \\
  \end{array}
\right)\,,
\quad
M_2 \equiv
\left(
  \begin{array}{ccc}
    0  &     0     & 1 \\
    0  & \;\;1\;\; & 0 \\
    -1 &     0     & 0 \\
  \end{array}
\right)\,,
\quad
M_3 \equiv
\left(
  \begin{array}{ccc}
    0  & \;\;1\;\; & 0 \\
    -1 &     0     & 0 \\
    0  &     0     & 1 \\
  \end{array}
\right)\,.
\end{equation}
One such equivalent Higgs vacuum is
\begin{equation}
\Phi= M_2 \cdot \left(
          \begin{array}{c}
            0 \\
            0 \\
            \eta \\
          \end{array}
        \right)
=
\left(
          \begin{array}{c}
            \eta \\
            0 \\
            0 \\
          \end{array}
        \right)\,,
\end{equation}
which will be used for the $\widehat{S}$ \textit{Ansatz} later on.

%%\newpage%%tmp
\subsection{Extended $SU(3)$ YMH theory}
\label{subsec:QCD-type-theory}

The second $SU(3)$ \YMHth~considered has three triplets
of complex scalar fields,
$\Phi_\alpha$ for $\alpha=1,\,2,\,3$.
The classical action is given by
\begin{eqnarray}\label{eq:actionYMH2}
S &=& \int_{\bR^4}\; d^4x\;
\Bigg\{
\frachalf \, \tr\, F_{\mu\nu}  F^{\mu\nu} +
       \sum_{\alpha=1}^{3}
       \left[(D_\mu  \Phi_\alpha)^\dagger \,(D^\mu \Phi_\alpha)     -
       \lambda \, \left( \Phi_\alpha^\dagger \Phi_\alpha - \eta^2 \right)^2
       \right] %%\right.
\nonumber\\[1mm]
&&   %%\left.
- \lambda \, (\Phi_1^\dagger \Phi_2)\,(\Phi_2^\dagger \Phi_1)
- \lambda \, (\Phi_1^\dagger \Phi_3)\,(\Phi_3^\dagger \Phi_1)
- \lambda \, (\Phi_2^\dagger \Phi_3)\,(\Phi_3^\dagger \Phi_2)
\Bigg\}\,.
\end{eqnarray}
The Higgs fields have a global $U(1)\times U(1)\times U(1)$ symmetry.

The scalar vacuum fields can be chosen as
\begin{equation}\label{eq:actionYMH2-Higgs-vac}
\Phi_1= \left(
          \begin{array}{c}
            \eta \\
            0 \\
            0 \\
          \end{array}
        \right)\,,
\quad
\Phi_2= \left(
          \begin{array}{c}
            0 \\
            \eta \\
            0 \\
          \end{array}
        \right)\,,
\quad
\Phi_3= \left(
          \begin{array}{c}
            0 \\
            0 \\
            \eta \\
          \end{array}
        \right)\,,
\end{equation}
which give an equal mass
($m_A = g\,\eta$) to all eight gauge fields $ A_\mu^a$.
There are ten physical scalar modes ($3\times 3\times 2 - 8 =10$),
nine of which are massive for quartic Higgs coupling $\lambda >0$
and one of which remains massless.
This last massless mode can get a mass from a more complicated
Higgs sector, but, in this paper, we keep the relatively simple
extended $SU(3)$ YMH theory as given by \eqref{eq:actionYMH2}.

%%\newpage%%tmp
\section{$\widehat{S}$ Ansatz in the basic $SU(3)$ YMH theory}
\label{sec:Shat-Ansatz}

The logic behind the existence of the new sphaleron $\widehat{S}$
in $SU(3)$ \YMH~theory with a single Higgs triplet and
the derivation of the $\widehat{S}$ \textit{Ansatz} have been
explained in Ref.~\cite{KlinkhamerRupp2005}, but will
be briefly recalled below. For our present purpose,
the focus will be on the \textit{Ansatz} fields and the corresponding
energy density. Both will be specialized to the radial gauge.
Standard spherical polar coordinates $(r,\,\theta,\,\phi)$ are used,
defined, in terms of the Cartesian coordinates by
$(x,\,y,\,z)=(r\sin\theta\cos\phi,\,r\sin\theta\sin\phi,\,r\cos\theta)$.

\subsection{Minimax procedure}
\label{subsec:Minimax-procedure}

For completeness, we sketch how the \emph{Ansatz}
for $\widehat{S}$ was obtained in Ref.~\cite{KlinkhamerRupp2005}.
The idea is to consider the mathematical  space of finite-energy
gauge and Higgs field configurations of the theory considered.
A noncontractible 3-sphere
can be constructed in this configuration space,
where the 3-sphere is parameterized by spherical coordinates
with polar angles $\psi$ and $\mu$ and azimuthal angle $\alpha$.
One point $V$ of that 3-sphere (at $\psi=0$) corresponds to the
configurations of the vacuum.

Next, evaluate the energy for all configurations
of this noncontractible sphere~(NCS).
The point $V$ (at $\psi=0$) has energy $E=0$ and the
other points of the NCS
have $E>0$. The configuration at $\psi=\pi$
has extra discrete symmetries of the fields
and is, generically, the one with the highest energy.
The qualitative picture is that of a 3-sphere with
the lowest-energy point at $\psi=0$ and
the highest-energy point at $\psi=\pi$.

We now follow a minimax procedure:
the maximum configuration ($\psi=\pi$)
is minimized by improving the profile functions of the fields,
in order to arrive at a genuine solution ($\widehat{S}$) of the
YMH field equations
(which needs to be verified explicitly).
The same minimax procedure
for a noncontractible loop (1-sphere) has given the
sphaleron $S$~\cite{KlinkhamerManton1984}
and for a noncontractible 2-sphere has given the sphaleron
$S^\ast$~\cite{Klinkhamer1993};
see Sec. IV of Ref.~\cite{KlinkhamerRupp2003}
for a review and further references.

Details of the NCS for $\widehat{S}$
can be found in Ref.~\cite{KlinkhamerRupp2005}
and in Appendix~\ref{app:NCS-in-extended-SU3-YMHth} here,
where the two extra Higgs triplets can be neglected
for the NCS relevant to the basic $SU(3)$ YMH theory.

%%\newpage%%tmp
\subsection{Gauge and Higgs field Ans\"{a}tze}
\label{subsec:Gauge-and-Higgs-fields}

The $\widehat{S}$ gauge fields in the radial gauge
are given by~\cite{KlinkhamerRupp2005}
\begin{subequations}\label{AsphericalAnsatz}
\begin{eqnarray}
g\, \widehat{A}_0(r,\theta,\phi) &=& 0 \,,\\[4mm]
g\,\widehat{A}_\phi(r,\theta,\phi) &=&
\alpha_1(r,\,\theta) \,\cos\theta\; T_\rho +
\alpha_2(r,\,\theta) \;V_\rho +
\alpha_3(r,\,\theta) \,\cos\theta\;U_\rho + \no\\[2mm]
&&
\alpha_4(r,\,\theta) \;\frac{\lambda_3}{2 i } +
\alpha_5(r,\,\theta) \;\frac{\lambda_8}{2 i }  \,,\\[4mm]
g\, \widehat{A}_\theta(r,\theta,\phi) &=&
\alpha_6(r,\,\theta) \;T_\phi +
\alpha_7(r,\,\theta) \,\cos\theta\;V_\phi +
\alpha_8(r,\,\theta) \;U_\phi \,,\\[4mm]
g\, \widehat{A}_r(r,\theta,\phi) &=&  0 \,,
\end{eqnarray}
\end{subequations}
with real functions $\alpha_{i}(r,\,\theta)$ that
are required to have positive parity
with respect to reflection of the $z$-coordinate,
\begin{equation}
\alpha_{i}(r,\pi-\theta) = +\alpha_{i}(r,\,\theta)\,,
\quad \mathrm{for}\;\; i=1,\, \ldots,\, 8\, .
\label{eq:alpha-i-parity}
\end{equation}
The gauge fields \eqref{AsphericalAnsatz}
involve the following generators of
the $\mathsf{su(3)}$ Lie algebra:
\begin{subequations}\label{eq:TVUmatrices}
\begin{align}
\label{eq:Tmatrices}
T_\phi &\equiv
-\sin\phi\,\frac{\lambda_1}{2 i }+\cos\phi\,\frac{\lambda_2}{2 i }\,,
&
T_\rho & \equiv \cos\phi\, \frac{\lambda_1}{2 i } + \sin\phi\,
\frac{\lambda_2}{2 i } \,,
\nonumber \\[2ex]
T_3 &\equiv \frac{\lambda_3}{2 i }\,,
\end{align}
\begin{align}
\label{eq:Vmatrices}
V_\phi &\equiv +\sin\phi\,  \frac{\lambda_4}{2 i } +\cos\phi\,
\frac{\lambda_5}{2 i }\,,
&
V_\rho &\equiv \cos\phi\, \frac{\lambda_4}{2 i }  - \sin\phi\,
\frac{\lambda_5}{2 i } \,,
\nonumber \\[2ex]
V_3 &\equiv \frac{\sqrt{3}\,\lambda_8+\lambda_3}{4 i }\,,
\end{align}
\begin{align}
\label{eq:Umatrices}
U_\phi &\equiv
\sin(2\phi)\,\frac{\lambda_6}{2 i }+\cos(2\phi)\,\frac{\lambda_7}{2 i }\,,
&
U_\rho &\equiv \cos(2\phi)\,\frac{\lambda_6}{2 i }  - \sin(2\phi)\,
\frac{\lambda_7}{2 i } \,,
\nonumber \\[2ex]
U_3 &\equiv \frac{\sqrt{3}\,\lambda_8-\lambda_3}{4 i }\,,
\end{align}
\end{subequations}
which have the property
\begin{equation}
\partial_\phi X = [ -2 U_3, X]\,,
\end{equation}
with $X$ standing for any of the matrices defined in
Eqs.~\eqref{eq:Tmatrices}--\eqref{eq:Umatrices}.

The axial \textit{Ansatz} functions $\alpha_{i}(r,\,\theta)$
have the following boundary conditions at the coordinate origin ($r=0$):
\begin{equation}
\alpha_{i}(0,\,\theta) = 0\,,\quad \mathrm{for}\;\; i=1, \ldots, 8,
\label{alphaBCSorigin}
\end{equation}
on the symmetry axis ($\bar{\theta}=0,\pi$):
\begin{subequations} \label{alphaBCSaxis}
\begin{align}
\label{alphaBCSaxis12}
\alpha_{i}(r,\bar{\theta})&=
 \widetilde{\alpha}_{i}(r)\,  \sin\theta \,\bigr|_{\,\theta=\bar{\theta}}\,,
 &\mathrm{for}\;\; i&=1, 2\,,
\\[2mm]
\label{alphaBCSaxis345}
\alpha_{i}(r,\bar{\theta})&=
 \widetilde{\alpha}_{i}(r)\, \sin^2\theta \,\bigr|_{\,\theta=\bar{\theta}}\,,
&\mathrm{for}\;\; i&=3,4,5,
\\[2mm]
\label{alphaBCSaxis67}
\alpha_{i}(r,\bar{\theta})&=
 (-)^{i-5}\,\cos\theta\,\partial_\theta\,
\alpha_{i-5}(r,\,\theta)\,\bigr|_{\,\theta=\bar{\theta}}\,,
&\mathrm{for}\;\; i&=6,7,
\\[2mm]
\label{alphaBCSaxis8}
\alpha_{i}(r,\bar{\theta})&=
 \frachalf\,\cos\theta\,\partial_\theta\,
\alpha_{i-5}(r,\,\theta)\,\bigr|_{\,\theta=\bar{\theta}}\,,
&\mathrm{for}\;\; i&=8,
\end{align}
\end{subequations}
and towards spatial infinity:
\begin{equation}
\lim_{r \to \infty}  \,
\left(
\begin{array}{c}
\alpha_1(r,\,\theta)\\
\alpha_2(r,\,\theta)\\
\alpha_3(r,\,\theta)\\
\alpha_4(r,\,\theta)\\
\alpha_5(r,\,\theta)\\
\alpha_6(r,\,\theta)\\
\alpha_7(r,\,\theta)\\
\alpha_8(r,\,\theta)
\end{array}
\right)
=
\left(
\begin{array}{c}
-2\,\sin\theta\,(1+\sin^2\theta)
\\
2\,\sin\theta\,\cos^2\theta\\
- 2\sin^2\theta\\
-\sin^2\theta\,(1+2\,\sin^2\theta)
\\
\sqrt{3}\,\sin^2\theta\\
2\\
2\\
-2\,\sin\theta
\end{array}
\right)\,.
\label{alphaBCSinfinity}
\end{equation}

The $\widehat{S}$ Higgs fields are given by~\cite{KlinkhamerRupp2005}
\begin{eqnarray}
\widehat{\Phi}(r,\theta,\phi)
&=& \eta\, \bigl[\, \beta_1(r,\,\theta)\,\lambda_3
+\beta_2(r,\,\theta)\,\cos\theta\; \,2 i \,T_\rho
+\beta_3(r,\,\theta)\,2 i \,V_\rho \,\bigr]\, \vect{1\\0\\0} \no\\
&=& \eta\,\vectleft{
\beta_1(r,\,\theta)\\
\beta_2(r,\,\theta)\,\cos\theta\; e^{ i \phi} \\
\beta_3(r,\,\theta)\;e^{- i \phi}}\,,
\label{PhiAnsatz}
\end{eqnarray}
with real functions $\beta_{j}(r,\,\theta)$ that are
even under reflection of the $z$-coordinate,
\begin{align}
\beta_{j}(r,\pi-\theta) &= +\beta_{j}(r,\,\theta)\,,
\quad \mathrm{for}\;\; j=1,2,3.
\label{eq:beta-j-parity}
\end{align}

%%%%%\newpage%%tmp
The axial \textit{Ansatz} functions $\beta_{j}(r,\,\theta)$ have the
following boundary conditions at the coordinate origin ($r=0$):
\begin{equation}
\beta_1(0,\,\theta) =
\beta_2(0,\,\theta) =
\beta_3(0,\,\theta) = 0\,,
\label{betaBCSorigin}
\end{equation}
on the symmetry axis ($\bar{\theta}=0,\pi$):
\begin{subequations} \label{betaBCSaxis}
\begin{eqnarray}
\label{betaBCSaxis1}
\partial_\theta\,\beta_1(r,\,\theta)\,\bigr|_{\,\theta=\bar{\theta}}&=&0\,,
\\[2mm]
\label{betaBCSaxis23}
\beta_{j}(r,\bar{\theta})&=&
\widetilde{\beta}_{j}(r) \,\sin\theta\,\bigr|_{\,\theta=\bar{\theta}}\,,
\quad \mathrm{for}\;\; j=2,3,
\end{eqnarray}
\end{subequations}
and towards spatial infinity:
\begin{align}
\lim_{r \to \infty}  \,
\left(
\begin{array}{c}
\beta_1(r,\,\theta)\\
\beta_2(r,\,\theta)\\
\beta_3(r,\,\theta)
\end{array}
\right)
=
\left(
\begin{array}{c}
\cos^2\theta\\
-\sin\theta\\
-\sin\theta
\end{array}
\right)\,.
\label{betaBCSinfinity}
\end{align}
Note that boundary condition \eqref{betaBCSorigin} is tighter than
the one given in Ref.~\cite{KlinkhamerRupp2005}, which has only
$\partial_\theta\, \beta_1(0,\,\theta) =0$. The
boundary conditions \eqref{betaBCSorigin} give a vanishing
Higgs field at the origin, $\widehat{\Phi}(0,\,\theta) = 0$,
which is needed for the existence of fermion zero modes
if the theory \eqref{eq:actionYMH1} has additional
Weyl fermions with Yukawa couplings to the Higgs
(cf. Sec.~V of the review article~\cite{KlinkhamerRupp2003}).
Recall that appropriate fermion zero modes give rise to the non-Abelian
chiral gauge anomaly~\cite{Bardeen1969} as discussed in
Refs.~\cite{Klinkhamer1998,KlinkhamerRupp2005}.

To summarize, the radial-gauge \textit{Ansatz} for $\widehat{S}$
in the basic YMH theory involves 11 axial functions,
8 functions $\alpha_{i}(r,\,\theta)$ for the Yang--Mills gauge fields and
3 functions $\beta_{j}(r,\,\theta)$ for the Higgs fields.
The boundary conditions on $\alpha_{i}$  and $\beta_{j}$
at spatial infinity
make for vacuum-type fields with vanishing energy density
and those at the coordinate origin and on the symmetry axis
make for a finite energy density
(see also Sec.~\ref{subsec:analytic-solution-near-the-origin}).

%%\newpage%%tmp
\subsection{Energy functional}
\label{subsec:Energy- functional}

The energy functional of the YMH theory \eqref{eq:actionYMH1} is given by
\begin{equation}
E[A,\,\Phi] = \int_{\bR^3} \dvx \left[\, -\frachalf \, {\rm tr} (F_{mn})^2 +
  |D_m\Phi|^2 +\lambda \left( |\Phi|^2-\eta^2\right)^2 \,\right]
  \,,
\label{energyfunctional}
\end{equation}
where the spatial indices $m,n$ run over $1,2,3$.
The $\widehat{S}$ \emph{Ans\"{a}tze}
\eqref{AsphericalAnsatz} and \eqref{PhiAnsatz} then give
\begin{equation}
E\left[\widehat{A},\,\widehat{\Phi}\right] =
4\pi \int_{0}^{\infty} \dint{r} \int_{0}^{\pi/2}\dint{\theta}
\, r^2\sin\theta\;\, \widehat{e}(r,\,\theta) \,,
\label{Eansatz}
\end{equation}
where the energy density $\widehat{e}(r,\,\theta)$ contains contributions
from the Yang--Mills term, the kinetic Higgs term, and the Higgs potential
term in the energy functional,
\begin{align}
\widehat{e}(r,\,\theta) &=
\widehat{e}_{\rm \,YM}(r,\,\theta) + \widehat{e}_{\rm \,Hkin}(r,\,\theta)
+ \widehat{e}_{\rm  \,Hpot}(r,\,\theta)\,.
\label{edens}
\end{align}
This energy density is given
in Appendix~\ref{app:Shat-energy-density-basic-YMHth} and
turns out to be well-behaved due to the boundary conditions
on the axial \emph{Ansatz} functions $\alpha_{i}(r,\,\theta)$  and
$\beta_{j}(r,\,\theta)$.
The energy density has, moreover, a reflection symmetry,
\begin{equation}
\widehat{e}(r,\,\theta)=\widehat{e}(r,\,\pi-\theta) \,,
\end{equation}
which allows the range of $\theta$ in (\ref{Eansatz}) to be
restricted to $[0,\pi/2]$.

%%\newpage%%tmp
\section{Field equations and analytical results}
\label{sec:Field-equations-analytical-results}

\subsection{Reduced field equations}
\label{subsec:Reduced-field-equations}

As shown in Ref.~\cite{KlinkhamerRupp2005},
and verified independently for the present article,
the YMH field equations with $\widehat{S}$ \textit{Ansatz} fields inserted
reduce to the variational equations obtained from
the \textit{Ansatz} energy functional \eqref{Eansatz}.
In short, the $\widehat{S}$ \textit{Ansatz} is self-consistent.

The variational equations (partial differential equations)
from the \textit{Ansatz} energy functional \eqref{Eansatz}
are rather cumbersome and will not be given here
(all the necessary information is contained in the energy density
as given by Appendix~\ref{app:Shat-energy-density-basic-YMHth}).

\subsection{Analytic solution near the origin}
\label{subsec:analytic-solution-near-the-origin}
%%based on "More on the $SU(3)$ sphaleron.." (March 4, 2017  v0.46frk)

The variational equations of Sec.~\ref{subsec:Reduced-field-equations}
can be solved analytically near the origin ($r\sim 0$).
Making the radial coordinate $r$ dimensionless by multiplication with
$g v$,  the analytic solution of these partial differential
equations near the origin ($r\sim 0$)
gives the following \textit{Ansatz} functions:
\begin{subequations}\label{alphas-betas-origin}
\begin{eqnarray}
\left(
\begin{array}{c}
\alpha_1(r,\,\theta)\\
\alpha_2(r,\,\theta)\\
\alpha_3(r,\,\theta)\\
\alpha_4(r,\,\theta)\\
\alpha_5(r,\,\theta)\\
\alpha_6(r,\,\theta)\\
\cos^2\theta\,\alpha_7(r,\,\theta)\\
\alpha_8(r,\,\theta)
\end{array}
\right)
&\sim&
\left(
\begin{array}{c}
c_{1}\,r^{2}\,\sin\theta
\\
c_{2}\,r^{2}\,\sin\theta\,|\cos\theta|
\\
c_{3}\,r^{3}\,\sin^2\theta
\\
c_{4}\,r^{2}\,\sin^2\theta
\\
c_{5}\,r^{2}\,\sin^2\theta
\\
-c_{1}\,r^{2}\,
\\
c_{2}\,r^{2}\,|\cos\theta|\,
\\
c_{3}\,r^{3}\,\sin\theta
\end{array}
\right)\,,
\label{alphas-origin}
%\end{eqnarray}
%\begin{eqnarray}
\\[2mm]
\left(\begin{array}{c}
\beta_1(r,\,\theta)\\
\beta_2(r,\,\theta)\\
\beta_3(r,\,\theta)
\end{array}\right)
&\sim&
\left(
\begin{array}{c}
c_{6}\,r\,|\cos\theta|
\\
c_{7}\,r^{2}\,\sin\theta
\\
c_{8}\,r\,\sin\theta
\end{array}\right)\,,
\label{betas-origin}
\end{eqnarray}
\end{subequations}
with constants $c_{1}$, \dots, $c_{8}$.
The functions \eqref{alphas-betas-origin},
with nonzero constants $c_{k}$,
make that the energy density at the origin is finite (positive) and
regular (no $\theta$ dependence as $r\to 0$).

%%%%%%%%%%%%%\newpage%%tmp
At this moment, recall the behavior
of the \textit{Ansatz} functions
towards infinity ($r\to\infty$) as given by
\eqref{alphaBCSinfinity} and \eqref{betaBCSinfinity},
but consider the combination $\cos^2\theta\,\alpha_7(r,\,\theta)$
instead of $\alpha_7(r,\,\theta)$.
The remarkable observation is that
the qualitative $\theta$-behavior of these
\textit{Ansatz} functions [including the combination
$\cos^2\theta\,\alpha_7(r,\,\theta)$]
is similar towards the origin and towards infinity, provided
$\{c_{1},\,c_{3},\,c_{4},\,c_{7},\,c_{8}\}$ are taken negative and
$\{c_{2},\,c_{5},\,c_{6}\}$ positive.
%This observation can also be used
%to get an approximation of the $\widehat{S}$ fields.
This observation underlies the useful redefinition of
the \textit{Ansatz} functions employed in
Appendix~\ref{app:Expansion-coefficients-basic-YMHth}.

Equation \eqref{betas-origin} gives the following behavior
of the triplet Higgs field near the origin ($r\sim 0$):
\begin{eqnarray}
\Phi_{\widehat{S}}\,(x,\,y,\,z)
&\sim&
\eta \;
\left(
\begin{array}{c}
c_{6}\,|z|
\\
0   %%c_{7}\,z\,(x+iy)
\\
c_{8}\,(x-iy)
\end{array}\right)\,,
\label{Higgs-Shat-origin}
\end{eqnarray}
with dimensionless Cartesian coordinates and
the second component being $\text{O}(r^2)$ for $r^2 \equiv x^2+y^2+z^2$.
The Higgs field \eqref{Higgs-Shat-origin}
shows a cusp-like behavior for the first component.
Still, the energy density involving
the Higgs field is well-behaved near the origin.
For comparison, the $SU(2)$ sphaleron $S$~\cite{KlinkhamerManton1984}
has the following behavior of the doublet Higgs field
near the origin (again with dimensionless Cartesian coordinates):
\begin{eqnarray}
\Phi_{S}(x,\,y,\,z)
&\sim&
\frac{v}{\sqrt{2}}\;c_{h}\;
\left(
\begin{array}{c}
x+iy
\\
z
\end{array}\right)\,,
\label{Higgs-S-origin}
\end{eqnarray}
which is perfectly smooth.

We can provide the following heuristic explanation of the
different behavior of the $S$ and $\widehat{S}$ Higgs
fields at the origin.
If the Higgs behavior near the origin is given by $\Phi \propto r$,
then $S$ gets a component $c_{h}\, z$
because the corresponding $\cos\theta$ behavior at infinity is odd under $\theta\to\pi-\theta$,
whereas $\widehat{S}$ gets a component $c_{6}\,|z|$
because the corresponding $\cos^2\theta$ behavior at infinity  is even.

%%\newpage%%tmp
\section{Numerical results}
\label{sec:Numerical-results}

\subsection{Minimization setup}
\label{subsec:Minimization-setup}

In order to apply numerical  minimization techniques,
we approximate the energy functional \eqref{Eansatz}
by an energy function of expansion coefficients,
where the relevant energy density \eqref{edens} has been detailed in
Appendix~\ref{app:Shat-energy-density-basic-YMHth}.
For this, we expand the two-dimensional profile functions
$\alpha_{i}(r,\,\theta)$ and $\beta_{j}(r,\,\theta)$
in nested orthogonal functions,
as done in previous work~\mbox{\cite{Haberichter2009,Schuh2014,Nagel2014}}
on the $\widehat{S}$ numerics.

For the radial expansion,
we switch to a compact radial coordinate $x$ defined by
\bsubeqs\label{eq:x-def}
\bea
  x &\equiv& \frac{g v r}{\chi + g v r} \in [0,\,1],
\\[2mm]
\chi&\in& \mathbb{R}^{+}\,,
\eea\esubeqs
with $v\equiv \sqrt{2}\,\eta$
as mentioned in Sec.~\ref{subsec:Embedded-electroweak-theory}.
The other coordinate, the polar angle $\theta$, is compact by definition and can be restricted to the following domain by use of the reflection symmetry:
\bea\label{eq:theta-domain}
\theta&\in& [0,\,\pi/2]\,.
\eea
The details of the
expansion coefficients for the \textit{Ansatz} functions
are relegated to Appendix~\ref{app:Expansion-coefficients-basic-YMHth}.

The double expansion in $x$ and $\theta$ of the \textit{Ansatz} functions
gives asymptotically ($M,\,N \rightarrow\infty$)
the following total number of coefficients from
\eqref{eq:N-coeff-basic-SU3-YMHth}:
\begin{equation}\label{eq:N-coeff-basic-SU3-YMHth-asymptotic}
  N_{\text{coeff}}^{(\text{basic\;YMHth})}
  \sim
  22\,N\,M\,.
\end{equation}
The asymptotic behavior \eqref{eq:N-coeff-basic-SU3-YMHth-asymptotic}
can be understood as follows:
$11$ \textit{Ansatz} functions (8 for the gauge fields and 3 for
the Higgs fields), a factor $(2\,N+1)\sim 2\,N$
from the $\theta$-expansion \eqref{eq:S_hat_angular_expansion_1_2},
and a factor  $(M+1)\sim M$
from the $x$-expansion \eqref{eq:S_hat_radial_expansion}.

%%\newpage%%tmp
\subsection{Numerical solution}
\label{subsec:Numerical-solution}

The \textit{Ansatz}-function expansions
presented in Appendix~\ref{app:Expansion-coefficients-basic-YMHth}
produce the $\widehat{S}$ energy as a function of the
expansion coefficients. The task, now, is to find the
optimal coefficients for an energy minimum
(recall that finding the perfect coefficients
corresponds to solving the reduced field equations).

As a first step,  we employ the simulated annealing (SA)
method~\cite{Kirkpatrick-etal1983},
a randomized global minimizer to give,
within a reasonable runtime, the best possible
set of initial values for the second step.
That second step is
a quadratically-convergent local minimizer based on
the Sequential Least-Squares Quadratic Programming (SLSQP)
method~\cite{Kraft1988}.

For our numerical calculations,
a C++ program of the first SA step has been written from the
ground up,
as an alternative to using one from the many available libraries.
The program of the second step relies upon
the SLSQP implementation of the Python library SciPy~\cite{SciPy2001}.

As the analytic integrations of the energy functional are
typically not feasible due to the size, the integrations over
$x$ and $\theta$ must be carried out numerically.
The numerical integrations over $x$ and $\theta$ are done
with the composite
Simpson's rule over a mesh given by the nodes of Chebyshev
polynomials of sufficiently large degree. This choice of grid
spacing is known to minimize the effect of Runge's phenomenon,
which occurs if the grid size does not exceed the expansion order by much.
[As a check of these numerical integrations, we have also
performed analytic integrations for relatively low expansion
orders, the largest being $(M,\,N)=(3,\,1)$
with some $2.3\times 10^6$ summands in the resulting
energy function.]

\vspace*{0cm}
\begin{figure*}[p] %[!h] %[p] %[!ht]
%\centering
\begin{center}  %%problem at arXiv, change to pdf figs??
\includegraphics[width=0.60\textwidth]{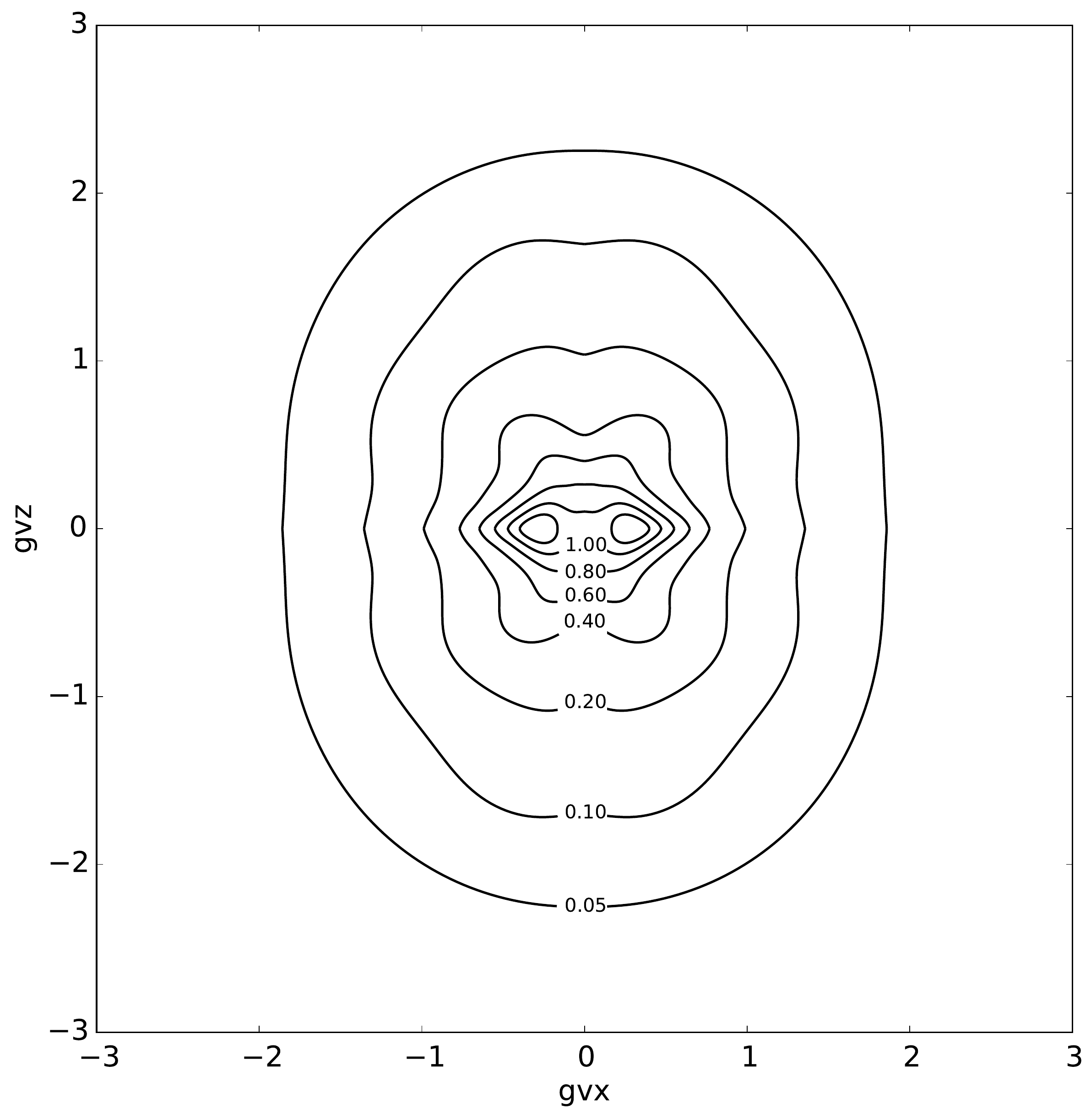}
%%fig12346-v204.eps  --> fig12346-v3.eps
%%fig5-v2040.eps  --> fig5-v3.eps
%%{Shat-num-sol-fig1-v2.eps}
%%{Shat-num-sol-fig-basicYMHth-edens-contours-M11-N3-24apr17.eps}
\end{center}
\vspace*{-2mm}
\caption{Energy density \eqref{edens} of the numerical $\widehat{S}$
solution in the basic $SU(3)$ \YMHth~\eqref{eq:actionYMH1} for $\lambda/g^2=0$
and $v\equiv \sqrt{2}\, \eta$.
The numerical solution is obtained from
minimization with expansion cutoffs $N=3$ and $M=18$,   %%v2
for a radial-compactification parameter $\chi=5$.
Cartesian coordinates $(x,\,y,\,z)$ are used and the plane $y=0$ is shown.
The two small contours around $(gvx,\,gvz)=(\pm 0.3,\,0)$
have the energy-density value $1.20$, in units of $(4\pi v/g)\,(gv)^3$.
}
\label{fig:energy-density-contours-basic-YMHth}
%\end{figure*}
\vspace*{2mm}
%\begin{figure*}[p] %[!h] %[p] %[!ht]
\centering
\includegraphics[width=0.60\textwidth]{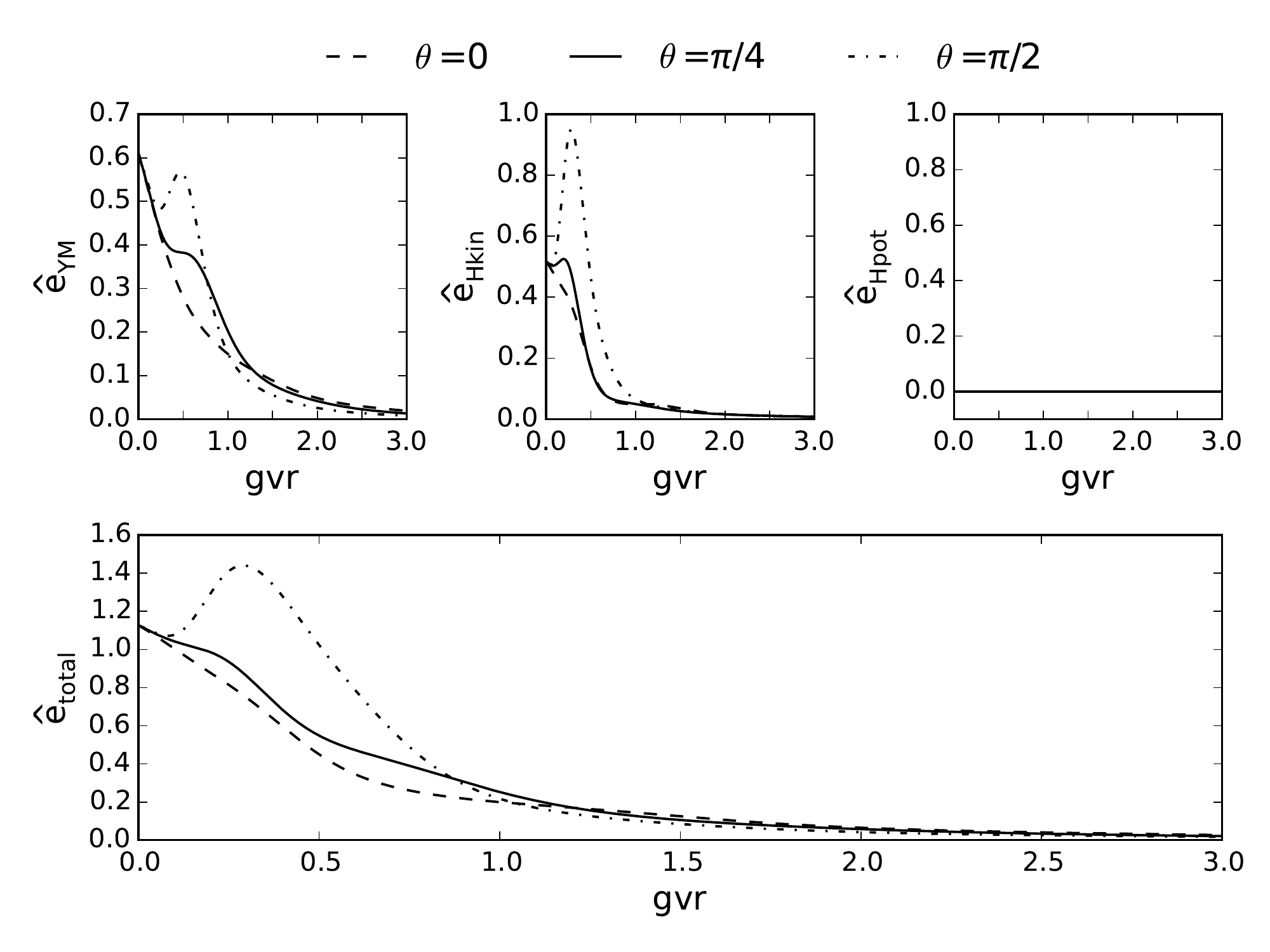}
%{Shat-num-sol-fig2-v1.eps}
%{Shat-num-sol-fig-basicYMHth-edens-slices-M11-N3-24apr17.eps}
\vspace*{-2mm}
\caption{Same as Fig.~\ref{fig:energy-density-contours-basic-YMHth},
but now for three slices at fixed polar angle $\theta$
and showing the various contributions to the total energy density, with
$\widehat{e}_{\rm  \,Hpot}=0$ for $\lambda/g^2=0$.
}
\label{fig:energy-density-slices-basic-YMHth}
\end{figure*}

For $\lambda/g^2=0$ and various expansion cutoffs $M$ and $N$, we find
the energies listed in Table~\ref{tab:E-Shat-num-sol-basic-SU3-YMHth}.
From this table, we obtain the following value of the
$\widehat{S}$ energy:
\begin{equation}\label{eq:EShat-num-basic-YMHth}
E_{\widehat{S}}^{(\text{basic\;$SU(3)$\;YMHth},\;\lambda/g^2=0)}
=
%(1.36 \pm 0.03)\times (4\pi v/g)\,,  %%v1
(1.35 \pm 0.03)\times (4\pi v/g)\,,  %%v2
\end{equation}
with a rough error estimate obtained from combining
the relative differences of energy
values in the last three rows of Table~\ref{tab:E-Shat-num-sol-basic-SU3-YMHth}
and the numerical relative error mentioned in the table caption.
For expansion cutoffs $M=18$ and $N=3$,
the energy densities are shown in
Figs.~\ref{fig:energy-density-contours-basic-YMHth}
and \ref{fig:energy-density-slices-basic-YMHth}.
The corresponding \textit{Ansatz} functions are not shown, as certain
gauge boson modes of the basic $SU(3)$ YMH theory are massless and
the convergence is slow.
[As the \textit{Ansatz} functions are not
perfectly converged, the contours of
Fig.~\ref{fig:energy-density-contours-basic-YMHth}
also need to be smoothed somewhat, especially
near the symmetry axis ($x=y=0$) and the equatorial plane ($z=0$).]
For the extended theory, all gauge boson modes  are massive and
the convergence is better
(see Sec.~\ref{subsec:Numerical-solution-in-extended-SU3-YMHth}).

\begin{table}[t]
\centering
\renewcommand{\tabcolsep}{1.1pc}    %% enlarge column spacing
\renewcommand{\arraystretch}{1.1}   %% enlarge line spacing
\begin{tabular}{cc|c}
  $M$  & $N$  & $E_{\widehat{S}}\,/\,\left[4\pi v/g\right]$  \\
   \hline\hline
%
%E(6,1) = 1.46768
%E(6,2) = 1.43247
%E(11,2) = 1.37102
%E(11,3) = 1.36038
%E(18,2) = 1.35310
%E(18,3) = 1.345
%
  $\phantom{0}3$ & $1$   & $1.610$ \\  %%v3
  $\phantom{0}6$ & $1$   & $1.468$ \\
  $\phantom{0}6$ & $2$   & $1.433$ \\
  $11$           & $2$   & $1.371$ \\
  $11$           & $3$   & $1.360$ \\
%  $18$           & $2$   & $1.353$   %%v1
  $18$           & $3$   & $1.345$    %%v2
\end{tabular}
\caption{Numerical estimates of the energy value of the sphaleron $\widehat{S}$
in the basic $SU(3)$ \YMHth~\eqref{eq:actionYMH1} with $\lambda/g^2=0$ and
definition $v\equiv \sqrt{2}\, \eta$.
The different energy values are obtained from numerical
minimization with various expansion cutoffs $M$ and $N$, where
the parameter $\chi=5$ is used for the radial compactification \eqref{eq:x-def}.
A conservative estimate of the numerical error on the energies quoted
is $10^{-2}\;4\pi v/g$.}
\vspace*{0cm}
\label{tab:E-Shat-num-sol-basic-SU3-YMHth}
\end{table}

The $\widehat{S}$ energy distribution of
Fig.~\ref{fig:energy-density-contours-basic-YMHth}
shows a nontrivial core ($gvr \lesssim 0.75$),
but the suggested ring  structure (with center at $gvr \sim 0.3$
in the $\theta=\pi/2$ plane) needs to be confirmed by
further calculations.
Somewhat further out ($1 \lesssim gvr \lesssim 2$), and with respect to
the axial-symmetry axis (the $z$-axis in our coordinate system),,
the energy distribution is slightly
prolate (equatorial radius smaller than polar radius).
The main contribution to the total energy comes from $gvr \sim 4$.

%%\newpage%%tmp
\subsection{Discussion}
\label{subsec:Discussion}

The result for the energy $E_{\widehat{S}}$
obtained in Sec.~\ref{subsec:Numerical-solution}
may be compared to the energy
$E_{S}$ of the embedded $SU(2)\times U(1)$ sphaleron $S$,
which has the following value
(cf. Table~1 of Ref.~\cite{KlinkhamerLaterveer1990}
and Fig.~1 of Ref.~\cite{KunzKleihausBrihaye1992}):
\begin{eqnarray}\label{eq:E-S}
E_{S}^{(\theta_w=\pi/6,\;\lambda/g^2=0)} &\approx&
1.505 \times (4\pi v/g)\,,
\end{eqnarray}
where we used $v\equiv \sqrt{2}\,\eta$ as mentioned
in Sec.~\ref{subsec:Embedded-electroweak-theory}.
With the numerical result \eqref{eq:EShat-num-basic-YMHth}
for the $\widehat{S}$ energy at $\lambda/g^2=0$, we then have
the following ratio:
\begin{equation}\label{eq:EShat-over-ES}
E_{\widehat{S}}^{(\lambda/g^2=0)}/E_{S}^{(\theta_w=\pi/6,\,\lambda/g^2=0)}
\approx
0.90\,,
\end{equation}
which is definitely below unity.
(Hints of an $E_{\widehat{S}}/E_{S}$ ratio
below unity were, first, reported in Ref.~\cite{Haberichter2009}
and, later, in Refs.~\cite{Schuh2014,Nagel2014}.
The behavior of the $\widehat{S}$ fields near the origin
was, however, not correct in these earlier numerical calculations.)

The result \eqref{eq:EShat-over-ES} is remarkable
in that the $\widehat{S}$ solution excites all eight gauge fields
and the $S$  solution only four. The low energy value of $\widehat{S}$
is, most likely, due to the fact that
the \textit{Ansatz} \eqref{AsphericalAnsatz}
has azimuthal and polar gauge fields which are
evenly distributed over the Lie algebra.

%%\newpage%%tmp
\section{$\widehat{S}$ in the extended $SU(3)$ YMH theory}
\label{sec:Shat-in-extended-SU3-YMHth}

The construction of $\widehat{S}$ in the extended $SU(3)$
\YMH~theory \eqref{eq:actionYMH2}
follows that of $\widehat{S}$ in the basic $SU(3)$
\YMH~theory \eqref{eq:actionYMH1}
as given in Ref.~\cite{KlinkhamerRupp2005}
and we can be relatively brief as regards the motivation of
the \textit{Ansatz}.
As explained in Sec.~\ref{subsec:Minimax-procedure},
the crucial element for the
$\widehat{S}$ \textit{Ansatz} is a noncontractible sphere
of configurations, which, for the extended $SU(3)$
\YMH~theory, is presented in Appendix~\ref{app:NCS-in-extended-SU3-YMHth}.

%%%%%%%%%%%%%%%%%%%%%%%\newpage%%tmp
\subsection{$\widehat{S}$ Ansatz}
\label{subsec:Shat-Ansatz-in-extended-SU3-YMHth}

The proper \textit{Ansatz} for $\widehat{S}$ in the
extended $SU(3)$ YMH theory \eqref{eq:actionYMH2} corresponds
to a generalization of the fields
\eqref{eq:NCS-in-extended-SU3-YMHth-approx-Shat}
at the ``top''  ($\psi=\pi$) of the noncontractible sphere
of configurations
constructed in Appendix~\ref{app:NCS-in-extended-SU3-YMHth}.

For the radial gauge, the \textit{Ansatz} gauge fields $\widehat{A}$
are again given by \eqref{AsphericalAnsatz}
and the \textit{Ansatz} Higgs fields $\widehat{\Phi}_\alpha$
correspond to appropriate generalizations of the fields in
Eqs.~\eqref{eq:NCS-in-extended-SU3-YMHth-approx-Shat-Phi1}%
--\eqref{eq:NCS-in-extended-SU3-YMHth-approx-Shat-Phi3}:
\begin{subequations}\label{eq:Shat-in-extended-SU3-YMHth-Ansatz-Higgs-fields}
\begin{eqnarray}
\label{eq:Shat-in-extended-SU3-YMHth-Ansatz-Higgs-fields-1}
\widehat{\Phi}_{1}(r,\theta,\phi)
&=&
\eta\,
\left(
  \begin{array}{c}
  \beta_1(r,\,\theta)\\
  \cos\theta\,\beta_2(r,\,\theta)\;e^{ i \phi}\\
  \beta_3(r,\,\theta)\;e^{- i \phi}
  \end{array}
\right)\,,
\\[2mm]
\label{eq:Shat-in-extended-SU3-YMHth-Ansatz-Higgs-fields-2}
\widehat{\Phi}_{2}(r,\theta,\phi)
&=&
\eta\,
\left(
  \begin{array}{c}
  \beta_4(r,\,\theta)\;e^{- i \phi}\\
  \cos\theta\,\beta_5(r,\,\theta)\\
  \beta_{6}(r,\,\theta)
  \end{array}
\right)\,,
\\[2mm]
\label{eq:Shat-in-extended-SU3-YMHth-Ansatz-Higgs-fields-3}
\widehat{\Phi}_{3}(r,\theta,\phi)
&=&
\eta\,
\left(
  \begin{array}{c}
  \cos\theta\,\beta_{7}(r,\,\theta)\; e^{ i \phi}\\
  \beta_{8}(r,\,\theta)\; e^{2 i \phi}\\
  \cos\theta\,\beta_{9}(r,\,\theta)
  \end{array}
\right)\,,
\end{eqnarray}
\end{subequations}
with real functions $\beta_{k}(r,\,\theta)$ that
are even under reflection of the $z$-coordinate,
\begin{align}\label{eq:Shat-in-extended-SU3-YMHth-Ansatz-Higgs-fields-betas-even}
\beta_{k}(r,\pi-\theta) &= +\beta_{k}(r,\,\theta)\,,
\quad\text{for}\;\;k=1,\,\ldots\,,9\,.
\end{align}
The \textit{Ansatz}
\eqref{eq:Shat-in-extended-SU3-YMHth-Ansatz-Higgs-fields-1}
for the first triplet is the same as
\eqref{PhiAnsatz} for the basic $SU(3)$ YMH theory.
In addition, there are the following boundary conditions
at the origin and toward infinity
\begin{subequations}\label{eq:Shat-in-extended-SU3-YMHth-Ansatz-Higgs-fields-bcs-origin-infty}
\begin{eqnarray}
\label{eq:Shat-in-extended-SU3-YMHth-Ansatz-Higgs-fields-bcs-origin}
\beta_{k}(0,\,\theta) &=& 0\,,
\quad\text{for}\;\;k=1,\,\ldots\,,9\,,
\\[2mm]
\label{eq:Shat-in-extended-SU3-YMHth-Ansatz-Higgs-fields-bcs-infty}
\lim_{r \to \infty}  \,
\left(
\begin{array}{c}
\beta_1(r,\,\theta)\\
\beta_2(r,\,\theta)\\
\beta_3(r,\,\theta)\\
\beta_4(r,\,\theta)\\
\beta_5(r,\,\theta)\\
\beta_{6}(r,\,\theta)\\
\beta_{7}(r,\,\theta)\\
\beta_{8}(r,\,\theta)\\
\beta_{9}(r,\,\theta)
\end{array}
\right)
&=&
\left(
\begin{array}{c}
\cos^2\theta\\
-\sin\theta\\
-\sin\theta\\
-\sin\theta\\
-1\\
0\\
-\sin\theta\\
\sin^2\theta\\
-1
\end{array}
\right)\,,
\end{eqnarray}
\end{subequations}
and the following boundary conditions
on the symmetry axis ($\bar{\theta}=0,\pi$):
\begin{subequations}
\label{eq:ext-YMHth-Ansatz-Higgs-bcs-axis}
\begin{eqnarray}
\label{eq:ext-YMHth-Ansatz-Higgs-bcs-axis1569}
\partial_\theta\,\beta_1(r,\,\theta)\,\bigr|_{\,\theta=\bar{\theta}}&=&0\,,
\quad \mathrm{for}\;\; k=1,\,5,\,6,\,9\,,
\\[2mm]
\label{eq:ext-YMHth-Ansatz-Higgs-bcs-axis2347}
\beta_{k}(r,\,\theta)\,\bigr|_{\,\theta=\bar{\theta}}&=&
\widetilde{\beta}_{k}(r) \,\sin\theta\,\bigr|_{\,\theta=\bar{\theta}}\,,
\quad \mathrm{for}\;\; k=2,\,3,\,4,\,7\,,
\\[2mm]
\label{eq:ext-YMHth-Ansatz-Higgs-bcs-axis8}
\beta_{8}(r,\,\theta)\,\bigr|_{\,\theta=\bar{\theta}}&=&
\widetilde{\beta}_{8}(r) \,\sin^2\theta\,\bigr|_{\,\theta=\bar{\theta}}\,.
\end{eqnarray}
\end{subequations}

To summarize, the radial-gauge \textit{Ansatz} for $\widehat{S}$
in the extended YMH theory involves 17 axial functions,
8 functions $\alpha_{i}(r,\,\theta)$ for the Yang--Mills gauge fields and
9 functions $\beta_{k}(r,\,\theta)$ for the Higgs fields.
Again, the boundary conditions on $\alpha_{i}$ and $\beta_{k}$
at spatial infinity
make for vacuum-type fields with vanishing energy density
and those at the coordinate origin and on the symmetry axis
make for a finite energy density.

%%\newpage%%tmp
\subsection{Analytic solution near the origin}
\label{subsec:analytic-solution-near-the-origin-in-extended-SU3-YMHth}

The energy density from the
$\widehat{S}$ \textit{Ans\"{a}tze} \eqref{AsphericalAnsatz}
and \eqref{eq:Shat-in-extended-SU3-YMHth-Ansatz-Higgs-fields}
in the extended YMH theory is given
in Appendix~\ref{app:Shat-energy-density-extended-SU3-YMHth}.
The corresponding  variational equations have the following solution
near the origin ($r\sim 0$):
\begin{equation}\label{eq:beta456789-origin}
\left(
\begin{array}{c}
\beta_4(r,\,\theta)\\
\beta_5(r,\,\theta)\\
\beta_{6}(r,\,\theta)\\
\beta_{7}(r,\,\theta)\\
\beta_{8}(r,\,\theta)\\
\beta_{9}(r,\,\theta)
\end{array}
\right)
\sim
\left(\begin{array}{c}
	-a_{9}\,   r\,   \sin\theta \\
	-a_{10}\,  r \\
	a_{11}\,  r\,   |\cos\theta | \\
	-a_{12}\,  r^2\, \sin\theta \\
	a_{13}\,  r^2\, \sin^2\theta \\
	-a_{14}\,  r
\end{array}\right)\,,
\end{equation}
where some suggestive minus signs have been inserted,
so that the qualitative $\theta$-behavior at the origin matches the behavior  \eqref{eq:Shat-in-extended-SU3-YMHth-Ansatz-Higgs-fields-bcs-infty}
at infinity. The solutions for the other
eleven \textit{Ansatz} functions near the origin
have already been given in
\eqref{alphas-betas-origin} .

%%\newpage%%tmp
\subsection{Numerical solution}
\label{subsec:Numerical-solution-in-extended-SU3-YMHth}

The numerical minimization of the $\widehat{S}$
energy in the extended YMH theory parallels the
calculation in the basic YMH theory
and is summarized in
Appendix~\ref{app:Minimization-setup-extended-SU3-YMHth}.
The double expansion in $x$ and $\theta$
of the \textit{Ansatz} functions
gives asymptotically ($M,\,N \rightarrow\infty$)
the following total number of coefficients from
\eqref{eq:N-coeff-extended-SU3-YMHth}:
\begin{equation}\label{eq:N-coeff-extended-SU3-YMHth-asymptotic}
  N_{\text{coeff}}^{(\text{ext.\;YMHth})}
  \sim
  34\,N\,M\,.
\end{equation}
The asymptotic behavior \eqref{eq:N-coeff-extended-SU3-YMHth-asymptotic}
can be understood as follows:
$17$ \textit{Ansatz} functions (8 for the gauge fields and 9 for
the Higgs fields),
 a factor $(2\,N+1)\sim 2\,N$ from the $\theta$-expansion,
and a factor $(M+1)\sim M$ from the $x$-expansion.

\begin{figure*}[p] %[!ht]
	\centering
\includegraphics[width=0.60\textwidth]{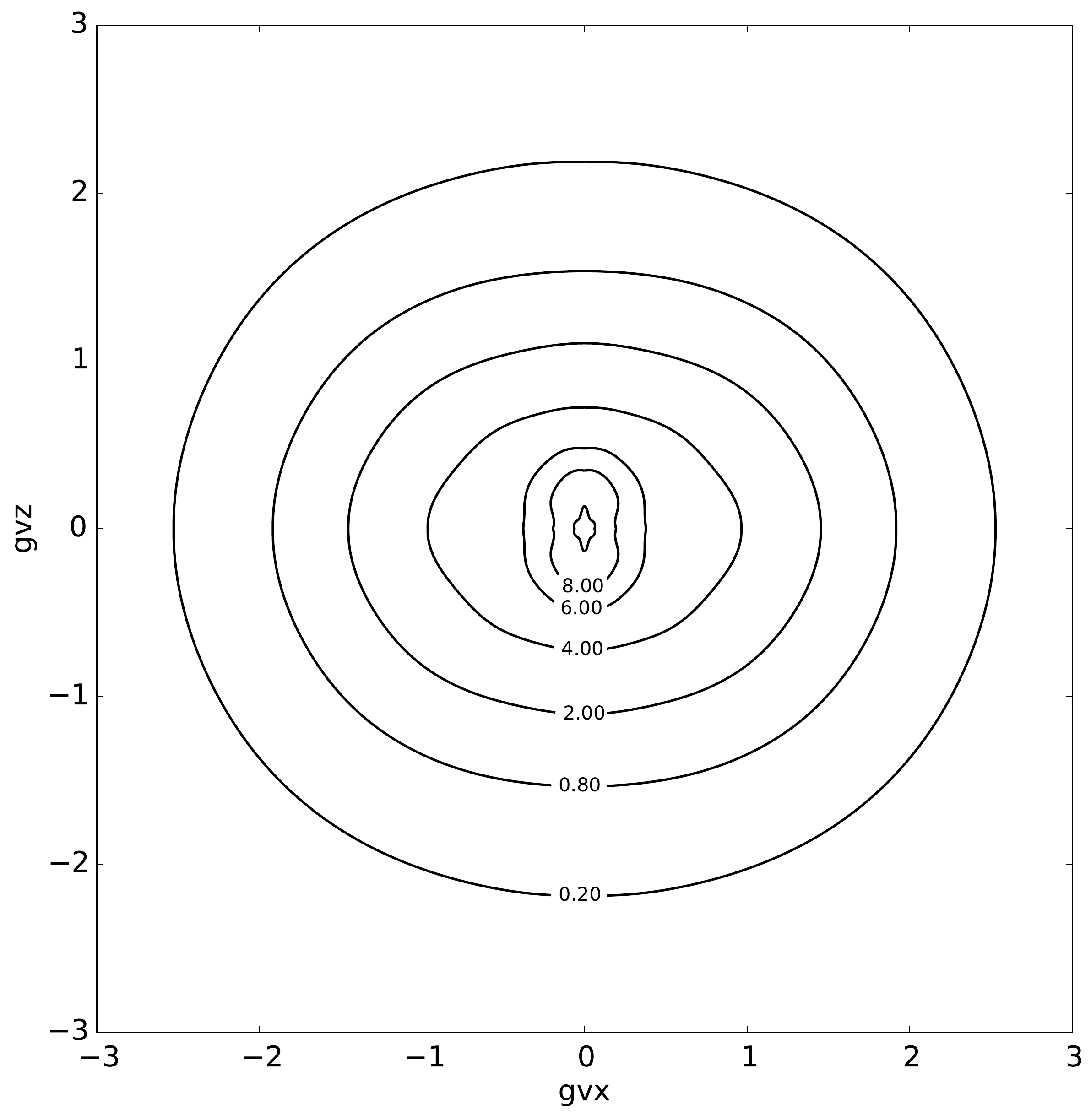}
%{Shat-num-sol-fig3-v1.eps}
%{Shat-num-sol-fig-extYMHth-edens-contours-M11-N3-24apr17.eps}
\vspace*{-2mm}
\caption{Energy density \eqref{eq:edens-in-extended-SU3-YMHth} of the numerical $\widehat{S}$
solution in the extended $SU(3)$ \YMHth~\eqref{eq:actionYMH2}
for $\lambda/g^2=1$ and $v\equiv \sqrt{2}\,\eta$.
The numerical solution is obtained from
minimization with expansion cutoffs $N=3$ and $M=11$,
for a radial-compactification parameter $\chi=3/2$.
Cartesian coordinates $(x,\,y,\,z)$ are used and the plane $y=0$ is shown.
The small contour around $(gvx,\,gvz)=(0,\,0)$
has the energy-density value $8.00$, in units of
$(4\pi v/g)\,(gv)^3$.
}
\label{fig:energy-density-contours-extended-YMHth}
%\end{figure*}
\vspace*{5mm}
%\begin{figure*}[!h] %[p] %[!ht]
	\centering
\includegraphics[width=0.60\textwidth]{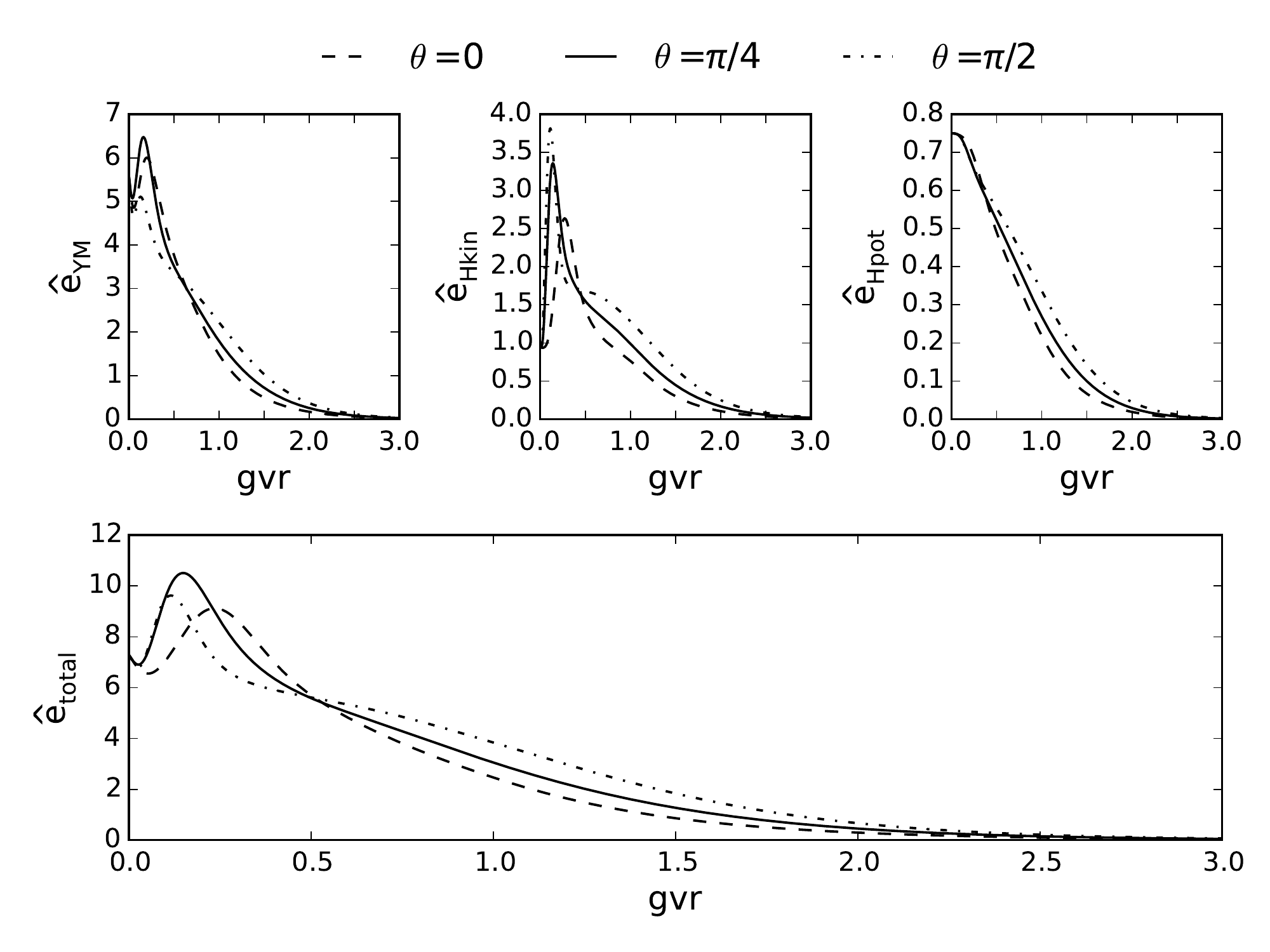}
%{Shat-num-sol-fig4-v1.eps}
%{Shat-num-sol-fig-extYMHth-edens-slices-M11-N3-24apr17.eps}
\vspace*{-2mm}
\caption{Same as Fig.~\ref{fig:energy-density-contours-extended-YMHth},
but now for three slices at fixed polar angle $\theta$
and showing the various contributions to the total energy density.
}
\label{fig:energy-density-slices-extended-YMHth}
\end{figure*}

For $\lambda/g^2=1$ and various expansion cutoffs $M$ and $N$,
we obtain the energies listed in Table~\ref{tab:energies_ext}.
From this table, we obtain the following value of the
$\widehat{S}$ energy:
\begin{equation}\label{eq:EShat-num-extended-YMHth}
E_{\widehat{S}}^{(\text{ext.\;$SU(3)$\;YMHth},\;\lambda/g^2=1)}
=
%(8.57 \pm 0.05)\times (4\pi \eta/g)\,,
(8.50 \pm 0.03)\times (4\pi \eta/g)\,,
\end{equation}
with a rough error estimate obtained from combining
the relative difference of energy
values in the last three rows of Table~\ref{tab:energies_ext}
and the numerical relative error mentioned in the table caption.
The various contributions to the total energy
have, for the $(M,\,N)=(11,\,3)$ numerical solution,
the approximate ratios
$E_\text{YM} : E_\text{Hkin} : E_\text{Hpot}$ $\approx$
%%$0.571 : 0.353 : 0.076$.  %%v1
$0.532:0.384:0.084$  %%v2
and the corresponding energy densities
are shown in Figs.~\ref{fig:energy-density-contours-extended-YMHth}
and \ref{fig:energy-density-slices-extended-YMHth}.
The main contribution to the total energy
comes from $gvr \sim 1.5$ (see Table~\ref{tab:energy_distribution}
for the build-up of the total energy).

Figures~\ref{fig:energy-density-contours-extended-YMHth}
and \ref{fig:energy-density-slices-extended-YMHth}
make clear that, with respect to
the axial-symmetry axis (the $z$-axis in our coordinate system),
the $\widehat{S}$ energy distribution for $gvr \gtrsim 0.6$
is slightly oblate (equatorial radius larger than polar radius),
whereas the energy distribution for $gvr \lesssim 0.4$
appears to be slightly prolate.

\begin{table}[t]
\centering
\renewcommand{\tabcolsep}{1.1pc}    %% enlarge column spacing
\renewcommand{\arraystretch}{1.1}   %% enlarge line spacing
\begin{tabular}{cc|c}
  $M$ & $N$ & $E_{\widehat{S}}\,/\,\left[4\pi \eta/g\right]$  \\
   \hline\hline
%
%E(3,1)  = 6.39033   %%v1: chi=5
%E(6,1)  = 6.09932
%E(6,2)  = 6.09055
%E(11,2) = 6.05968
%E(11,3) = 6.05817
%
%E(3,1)  = 8.627     %%v2: chi=1.5
%E(6,1)  = 8.527
%E(6,2)  = 8.526
%E(11,2) = 8.506
%E(11,3) = 8.503
%
  3 & 1
%  & $9.037$ %$=9.03729$  %$=\sqrt{2}\times 6.39033$ %%v1: chi=5
   & $8.627 $ %%v2: chi=1.5
  \\
  6 & 1
%  & $8.626$ %$=8.62574$ %$=\sqrt{2}\times 6.09932$ %%v1: chi=5
   &  $8.527$ %%v2: chi=1.5
  \\
  6 & 2
%  & $8.613$ %$=8.61334$ %$=\sqrt{2}\times 6.09055$ %%v1: chi=5
   & $8.526$ %%v2: chi=1.5
  \\
  11& 2
%  & $8.570$ %$=8.56968$ %$=\sqrt{2}\times 6.05968$ %%v1: chi=5
   & $8.506$ %%v2: chi=1.5
  \\
  11& 3
%  & $8.568$ %$=8.56755$ %$=\sqrt{2}\times 6.05817$  %%v1: chi=5
   & $8.503$ %%v2: chi=1.5
\end{tabular}
\caption{Numerical estimates of the energy value of the sphaleron $\widehat{S}$
in the extended $SU(3)$ \YMHth~\eqref{eq:actionYMH2} with $\lambda/g^2=1$
and $v\equiv \sqrt{2}\, \eta$.
The different energy values are obtained from numerical
minimization with various expansion cutoffs $M$ and $N$, where
the parameter $\chi=3/2$ is used for the radial compactification \eqref{eq:x-def}.
A conservative estimate of the numerical error on the energies quoted
is $10^{-2}\;4\pi \eta/g$.}
\label{tab:energies_ext}
\end{table}
\begin{table}[h!]
\centering
\renewcommand{\tabcolsep}{1.1pc}    %% enlarge column spacing
\renewcommand{\arraystretch}{1.1}   %% enlarge line spacing
\begin{tabular}{c|c}
  $gvR$ & $\widehat{E}_{R}/\widehat{E}_{\infty}$ \\
   \hline\hline
  0.3 & 0.0125 \\
  0.6 & 0.0719 \\
  0.9 & 0.1923 \\
  1.2 & 0.3574 \\
  1.5 & 0.5252 \\
  1.8 & 0.6785 \\
  2.1 & 0.7840 \\
  2.4 & 0.8649 \\
  2.7 & 0.9120 \\
  3.0 & 0.9449 \\
  4.0 & 0.9870 \\
  5.0 & 0.9958 \\
  6.0 & 0.9979 \\
\end{tabular}
\caption{Energy contribution up to radius $R$ of the numerical
$\widehat{S}$ solution
in the extended $SU(3)$ Yang--Mills--Higgs theory
\eqref{eq:actionYMH2} for
$\lambda/g^2=1$ and $v\equiv\sqrt{2}\,\eta$.
The numerical solution has been obtained from minimization with expansion cutoffs $N=3$ and $M=11$ using $\chi=3/2$.
The values of the partial energy
$\widehat{E}_{R} \equiv
4\pi\int_{0}^{R} dr \int_{0}^{\pi /2} d\theta\ r^2\sin\theta\ \widehat{e}(r,\,\theta)$
are given relative to the total %asymptotic
energy $\widehat{E}_{\infty}=E_{\widehat{S}}$ from Table~\ref{tab:energies_ext}.}
\label{tab:energy_distribution}
\end{table}

\vspace*{0cm}

In order to show the profile functions $\alpha_{i}(x,\,\theta)$
and $\beta_{k}(x,\,\theta)$ of the numerical solution,
we introduce the following rescalings with
angular functions:
\bsubeqs\label{eq:S_hat_alpha_beta_hats}
\bea\label{eq:S_hat_alpha_hats}
\left(\begin{array}{c}
  						\widehat{\alpha}_1(x,\,\theta) \\[2mm]
  						\widehat{\alpha}_2(x,\,\theta) \\[2mm]
  						\widehat{\alpha}_3(x,\,\theta) \\[2mm]
  						\widehat{\alpha}_4(x,\,\theta) \\[2mm]
  						\widehat{\alpha}_5(x,\,\theta) \\[2mm]
  						\widehat{\alpha}_6(x,\,\theta) \\[2mm]
  						\widehat{\alpha}_7(x,\,\theta) \\[2mm]
  						\widehat{\alpha}_8(x,\,\theta)
\end{array}\right)
  &=&	
\left(\begin{array}{c}
  						\alpha_1(x,\,\theta)/[-2\sin\theta(1+\sin^2\theta)] \\[2mm]
  						\alpha_2(x,\,\theta)/[2\sin\theta] \\[2mm]
  						\alpha_3(x,\,\theta)/[-2\sin^2\theta] \\[2mm]
  						\alpha_4(x,\,\theta)/[-\sin^2\theta(1+2\sin^2\theta)] \\[2mm]
  						\alpha_5(x,\,\theta)/[\sqrt{3}\sin^2\theta] \\[2mm]
  						\alpha_6(x,\,\theta)/2 \\[2mm]
  						\alpha_7(x,\,\theta)/2 \\[2mm]
  						\alpha_8(x,\,\theta)/[-2\sin\theta]
\end{array}\right),
%\\[4mm]
\eea
\bea
\label{eq:S_hat_beta_hats-basic+ext}
\left(\begin{array}{c}
  			\widehat{\beta}_1(x,\,\theta) \\[2mm]
  			\widehat{\beta}_2(x,\,\theta) \\[2mm]
  			\widehat{\beta}_3(x,\,\theta) \\[2mm]
            \widehat{\beta}_4(x,\,\theta) \\[2mm]
  			\widehat{\beta}_5(x,\,\theta) \\[2mm]
  			\widehat{\beta}_6(x,\,\theta) \\[2mm]
  			\widehat{\beta}_7(x,\,\theta) \\[2mm]
  			\widehat{\beta}_8(x,\,\theta) \\[2mm]
  			\widehat{\beta}_9(x,\,\theta)
\end{array}\right)
&=&
\left(\begin{array}{c}
  			\beta_1(x,\,\theta)\\[2mm]
  			\beta_2(x,\,\theta)/[-\sin\theta] \\[2mm]
  			\beta_3(x,\,\theta)/[-\sin\theta] \\[2mm]
            \beta_4(x,\,\theta)/[-\sin\theta] \\[2mm]
  			-\beta_5(x,\,\theta) \\[2mm]
  			\beta_6(x,\,\theta) \\[2mm]
  			\beta_7(x,\,\theta)/[-\sin\theta] \\[2mm]
  			\beta_8(x,\,\theta)/[\sin^2\theta] \\[2mm]
  			-\beta_9(x,\,\theta)
\end{array}\right),
\eea\esubeqs
where the divisions by $\sin\theta$ or $\sin^2\theta$
are allowed by the boundary conditions on the symmetry axis,
as given by Eqs.~\eqref{alphaBCSaxis}
and \eqref{eq:ext-YMHth-Ansatz-Higgs-bcs-axis}.
For these redefined \textit{Ansatz} functions, the
behavior at spatial infinity is simplified, with values
in the range $[0,\,1]$,
\begin{subequations}\label{eq:alphahat-betahat-bcs-infty}
\begin{eqnarray}
\hspace*{-0mm}
\lim_{x\to 1}
\left(
\begin{array}{c}
\widehat{\alpha}_1(x,\,\theta)\\[2mm]
\widehat{\alpha}_2(x,\,\theta)\\[2mm]
\widehat{\alpha}_3(x,\,\theta)\\[2mm]
\widehat{\alpha}_4(x,\,\theta)\\[2mm]
\widehat{\alpha}_5(x,\,\theta)\\[2mm]
\widehat{\alpha}_6(x,\,\theta)\\[2mm]
\cos^2\theta\;\widehat{\alpha}_7(x,\,\theta)\\[2mm]
\widehat{\alpha}_8(x,\,\theta)
\end{array}
\right)
&=&
\left(
\begin{array}{c}
1
\\[2mm]
\cos^2\theta
\\[2mm]
1
\\[2mm]
1
\\[2mm]
1
\\[2mm]
1\\[2mm]
\cos^2\theta\\[2mm]
1
\end{array}
\right),
\label{eq:alphahat-bcs-infty}
%\end{eqnarray}
%\begin{eqnarray}
\\[4mm]
\hspace*{-0mm}
\lim_{x\to 1}
\left(\begin{array}{c}
\widehat{\beta}_1(x,\,\theta)\\[2mm]
\widehat{\beta}_2(x,\,\theta)\\[2mm]
\widehat{\beta}_3(x,\,\theta)\\[2mm]
\widehat{\beta}_4(x,\,\theta)\\[2mm]
\widehat{\beta}_5(x,\,\theta)\\[2mm]
\widehat{\beta}_6(x,\,\theta)\\[2mm]
\widehat{\beta}_7(x,\,\theta)\\[2mm]
\widehat{\beta}_8(x,\,\theta)\\[2mm]
\widehat{\beta}_9(x,\,\theta)
\end{array}\right)
&=&
\left(
\begin{array}{c}
\cos^2\theta\\[2mm]
1\\[2mm]
1\\[2mm]
1\\[2mm]
1\\[2mm]
0\\[2mm]
1\\[2mm]
1\\[2mm]
1
\end{array}\right)\,.
\label{eq:betahat-bcs-infty}
\end{eqnarray}
\end{subequations}
\begin{figure*}[!ht]
	\centering
\includegraphics[width=.90\textwidth]{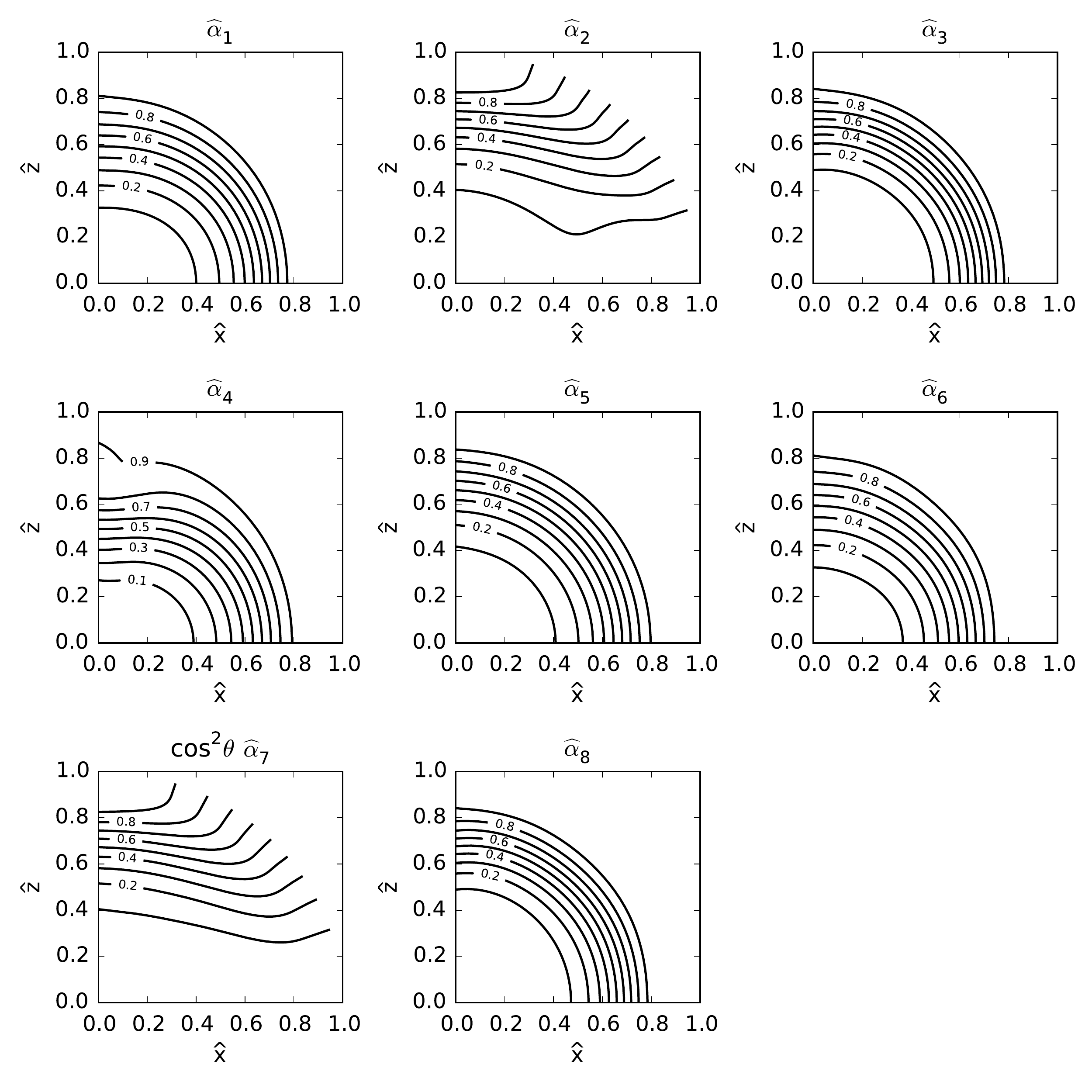}
%{Shat-num-sol-fig5-v1.eps}
%{Shat-num-sol-fig-extYMHth-alpha-widehat-M11-N3-24apr17.eps}
\caption{Equidistant contour plots of the rescaled profile
functions $\widehat{\alpha}_{i}(x,\,\theta)$ of the $(M,\,N)=(11,3)$
configuration obtained from numerical minimization
of the $\widehat{S}$ energy functional in the
extended Yang-Mills-Higgs theory \eqref{eq:actionYMH2}
for \mbox{$\lambda/g^2=1$}.
The rescaled profile functions $\widehat{\alpha}_{i}(x,\,\theta)$
are defined by \eqref{eq:S_hat_alpha_hats}.
Compactified Cartesian coordinates
$(\widehat{x},\,\widehat{y},\,\widehat{z})$ are used
and the plane $\widehat{y}=0$ is displayed.
Specifically, the two coordinates shown are given by
\mbox{$\widehat{x}= gvr\,\sin\theta/$}\mbox{$(gvr+\chi)$} and
$\widehat{z}= gvr\,\cos\theta/(gvr+\chi)$.
The numerical minimization procedure and the contour plots
both use $\chi=3/2$.}
\label{fig:Shat_num_LO_alphas_ext}
\vspace*{10cm}
\end{figure*}
\begin{figure*}[!ht]
	\centering
\includegraphics[width=.90\textwidth]{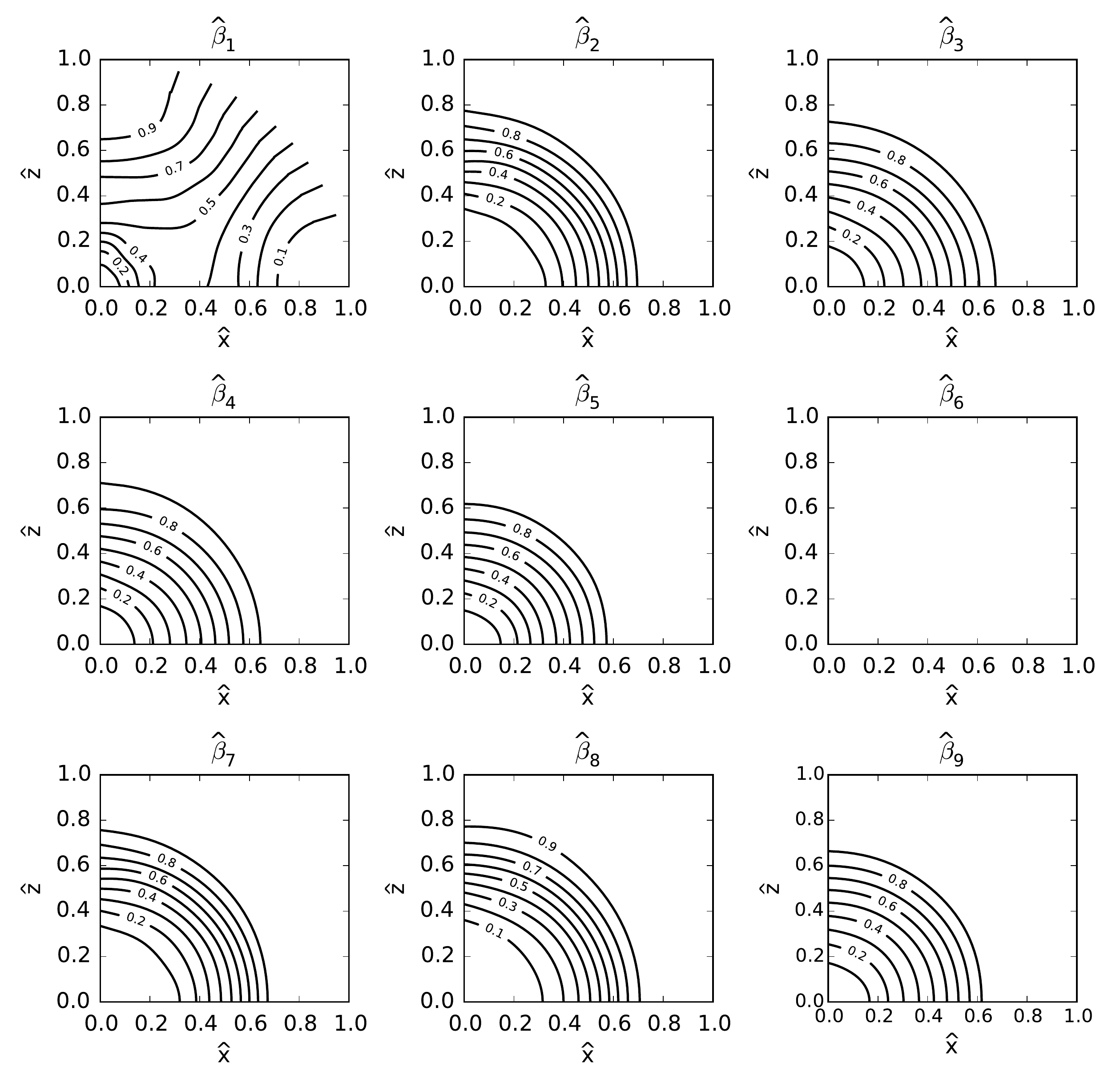}
%{Shat-num-sol-fig6-v1}
%{Shat-num-sol-fig-extYMHth-beta-widehat-M11-N3-24apr17.eps}
	\caption{Same as Fig.~\ref{fig:Shat_num_LO_alphas_ext},
but now with equidistant contour plots of the rescaled
profile functions $\widehat{\beta}_{k}(x,\,\theta)$
from \eqref{eq:S_hat_beta_hats-basic+ext}.
The numerical result for $\widehat{\beta}_{6}$
is close to zero,
$|\widehat{\beta}_{6}| \lesssim 3 \times 10^{-4}$.}
	\label{fig:Shat_num_LO_betas_ext}
\vspace*{10cm}
\end{figure*}
The boundary conditions at the origin match \eqref{alphaBCSorigin}
and \eqref{eq:Shat-in-extended-SU3-YMHth-Ansatz-Higgs-fields-bcs-origin}
of the original \textit{Ansatz} functions,
\bsubeqs\label{eq:alphahat-betahat-bcs-origin}
\bea
\widehat{\alpha}_{i}(0,\,\theta) &=& 0\,,
\quad \text{for}\;\; i=1,\, \ldots,\, 8\,,
\\[2mm]
\widehat{\beta}_{k}(0,\,\theta) &=& 0 \,,
\quad\text{for}\;\;  k=1,\, \ldots\,, 9\,.
\eea\esubeqs

Figures~\ref{fig:Shat_num_LO_alphas_ext} and \ref{fig:Shat_num_LO_betas_ext} 
present the rescaled profile functions of the numerical solution.
As mentioned in the caption of
Fig.~\ref{fig:Shat_num_LO_betas_ext},
the numerical solution for
$\widehat{\beta}_{6}(x,\,\theta)=\beta_{6}(x,\,\theta)$
is close to zero.
It can, indeed, be shown
that $\beta_{6}(x,\,\theta)=0$ solves the
$\beta_{6}$ variational equation from
\eqref{eq:edens23-in-extended-SU3-YMHth-Hkin2}
and \eqref{eq:edens23-in-extended-SU3-YMHth-Hpot123}.
With all Yang-Mills modes massive,
the energy densities and
profile functions appear to have converged reasonably well,
but the detailed behavior of
Figs.~\ref{fig:energy-density-contours-extended-YMHth}--%
\ref{fig:Shat_num_LO_betas_ext}
may still change somewhat with further minimization runs.

%%\newpage%%tmp
\subsection{Discussion}
\label{subsec:Discussion-in-extended-SU3-YMHth}

The $\widehat{S}$ gauge fields in the extended $SU(3)$ YMH theory
have a very special structure (as mentioned in the
last paragraph of Sec.~\ref{subsec:Discussion})
and we conjecture that these gauge fields may somehow play a role
in the nonperturbative dynamics of QCD.
It is true that the Higgs fields are important for
obtaining an equilibrium solution
($E_\text{YM}$ scales as $1/\overline{R}$
and $E_\text{Hkin}$ scales as $\overline{R}$,
with $\overline{R}$ the typical scale of the configuration).
In QCD, there are no such fundamental Higgs fields and it is not clear
how the $\widehat{S}$ gauge fields
would be prevented from expanding ($\overline{R}\to\infty$).
Still, it is not excluded that QCD quantum effects
produce attractive forces on this special lump of gauge fields.
In any case, it appears that the Yang-Mills configuration
space near the $\widehat{S}$ gauge field configuration
is relatively flat and this static three-dimensional
configuration may play a role in a Hamiltonian
analysis. (The corresponding instanton-type configuration
$\widehat{I}$
[which has NCS gauge fields \eqref{eq:NCS-in-extended-SU3-YMHth-A0}
and \eqref{eq:NCS-in-extended-SU3-YMHth-Am} with
$\psi=\psi(t)$ and, for example, $\mu=\alpha=\pi/2$]
may play a role in the Euclidean path integral).

The result for the energy $E_{\widehat{S}}$
obtained in Sec.~\ref{subsec:Numerical-solution-in-extended-SU3-YMHth}
can be compared to the following
nonperturbative ``soliton'' energy scale:
\begin{subequations}\label{eq:E-gl-soliton-m-gl-alpha-gl}
\begin{eqnarray}
\label{eq:E-gl-soliton}
E_\text{gl,\,soliton} &\equiv& m_\text{gl}/\alpha_\text{gl}\,,
\end{eqnarray}
defined in terms of the ``gluon mass'' and the
``gluon fine-structure constant,''
\begin{eqnarray}
\label{eq:m-gl}
m_\text{gl} &\equiv& g\,\eta\,,
\\[2mm]
\label{eq:alpha-gl}
\alpha_\text{gl} &\equiv& g^2/(4\pi)\,,
\end{eqnarray}
\end{subequations}
where the last two  right-hand-sides involve quantities of our classical
extended $SU(3)$ YMH theory \eqref{eq:actionYMH2}.
With the numerical result \eqref{eq:EShat-num-extended-YMHth}
for the $\widehat{S}$ energy, we then have the following ratio:
\begin{equation}\label{eq:EShat-over-Enonpert}
E_{\widehat{S}}/E_\text{gl,\,soliton}
\approx
8.5\,.
\end{equation}

Another characteristic of $\widehat{S}$ is its size.
Table~\ref{tab:energy_distribution} shows that
the radius for which the energy has reached $90\%$
of its asymptotic value is approximately
$2.60/(gv) \approx 1.84/(g\eta)$
and the corresponding diameter is then
\begin{equation}\label{eq:diameter-Shat}
d_{\widehat{S}}  \approx \frac{3.7}{m_\text{gl}} \,,
\end{equation}
where $m_\text{gl}$ has been defined by \eqref{eq:m-gl}.

With the cautionary remarks of the first paragraph of this subsection
in mind, we now turn to QCD and consider the $\widehat{S}$ gauge fields
obtained in Sec.~\ref{subsec:Numerical-solution-in-extended-SU3-YMHth}.
From QCD, we take over
$m_\text{gl} \sim (\text{fm})^{-1} \sim 200\;\text{MeV}$  and
$\alpha_\text{gl}\sim \alpha_s(200\;\text{MeV}) \sim 1$
(cf. Fig.~9.3 of Ref.~\cite{PDG2016}),
so that $E_\text{gl,\,soliton}\sim 200\;\text{MeV}$.
Then, ratio \eqref{eq:EShat-over-Enonpert} gives
$E_{\widehat{S}} \sim 8.5 \times 200\;\text{MeV} \sim 1.7 \;\text{GeV}$
in a QCD context. Similarly, the $\widehat{S}$  diameter \eqref{eq:diameter-Shat}
would correspond to $3.7\,\text{fm}$ in a QCD context and
Fig.~\ref{fig:energy-density-contours-extended-YMHth}
would give the energy-density contours
(scaled by a factor of $1/2$ perhaps)
for Cartesian coordinates $x$ and $z$ in units of $0.71\,\text{fm}$.
We conjecture that the $\widehat{S}$ gauge fields
(with an energy of order $0.8 \;\text{GeV}$ perhaps)
may contribute substantially to the field content of QCD glueballs
(cf. p.~798 of Ref.~\cite{PDG2016}).

Let us place
our suggestion about QCD glueballs in context.
It is, by now, well-known that, in an effective meson theory (motivated by
QCD with an infinitely large number $N_{c}$ of colors~\cite{tHooft1974}),
baryons may be considered as solitons~\cite{Skyrme1961,Witten1979,Witten1983}.
But there appears to be no place for glueballs
in this effective meson theory.
For this reason, we suggest to use the
extended $SU(3)$ YMH theory \eqref{eq:actionYMH2}
as a complementary effective theory,
without mesons and baryons, but possibly with
glueballs as solitons/sphalerons.
Admittedly, the extended $SU(3)$ YMH theory
would not have linear (flux-tube) confinement of gluons,
but the gauge bosons would be massive and not reach far out.
A more serious problem is the apparent lack of a small parameter
in QCD, which would support the use of semiclassical methods in the
effective YMH theory.

%%\newpage%%tmp
\section{Conclusion}
\label{sec:Conclusion}

In this article, we have obtained the numerical solutions of
the sphaleron $\widehat{S}$ in two $SU(3)$ \YMH~theories,
one with a single Higgs triplet and another with three Higgs triplets.
There were two crucial steps in getting these
numerical results. The
first step was that we managed
to obtain the respective analytic solutions of the
\textit{Ansatz} functions near the coordinate origin.
The second step was to use a mixed analytical-numerical procedure,
namely, to expand the \textit{Ansatz} functions
in orthogonal polynomials, to perform the
energy integrals analytically for low expansion orders
or numerically for larger expansion orders, and, finally, to
use an efficient numerical minimization procedure
over the expansion coefficients in the remaining expression for the
energy.

There are, at least, three outstanding issues. The first issue is
to numerically obtain
the corresponding fermion zero modes, based on the
\textit{Ans\"{a}tze} of Ref.~\cite{KlinkhamerRupp2005}.
The second issue is to perform the stability analysis
of the $\widehat{S}$ solutions found in the two $SU(3)$
\YMH~theories considered.
The third issue is, depending on the outcome of this stability
analysis ($\widehat{S}$ being unstable or perhaps metastable),
to determine the proper role of the $\widehat{S}$ gauge fields
in the nonperturbative dynamics of
quarkless
quantum chromodynamics.

%%%%\newpage%%tmp
\section*{\hspace*{-5mm}ACKNOWLEDGMENTS}
\vspace*{-0mm}\noindent
FRK thanks J. Greensite for useful discussions on QCD.

%%\newpage%%tmp
\begin{appendix}
\section{$\widehat{S}$ energy density in the basic $SU(3)$ YMH theory}
\label{app:Shat-energy-density-basic-YMHth}

In this appendix, we present the $\widehat{S}$
energy density \eqref{edens} of the radial-gauge \textit{Ansatz}
fields \eqref{AsphericalAnsatz}
and \eqref{PhiAnsatz} in the basic $SU(3)$ Yang--Mills--Higgs theory
\eqref{eq:actionYMH1}.
The following expressions are, in fact, equivalent to the energy densities
from Ref.~\cite{KlinkhamerRupp2005}
for $\alpha_9=\alpha_{10}=\alpha_{11}=0$:

\begin{align}
\label{eq:edens-YM}
  \widehat{e}_\textrm{\,YM}&=
  \frac{1}{2g^2r^2\sin^2\theta}\,
  \bigg\{ \cos^2\theta\left(\partial_r\alpha_1\right)^2
  +\left(\partial_r\alpha_2\right)^2 + \cos^2\theta\left(\partial_r\alpha_3\right)^2
  +\left(\partial_r\alpha_4\right)^2 +\left(\partial_r\alpha_5\right)^2\bigg\}
  \nonumber \\
  &+ \frac{1}{2g^2r^2}\,\bigg\{\left(\partial_r\alpha_6\right)^2
  + \cos^2\theta\left(\partial_r\alpha_7\right)^2
  + \left(\partial_r\alpha_8\right)^2\bigg\}
  \nonumber \\
  &+
  \frac{1}{2g^2r^4\sin^2\theta}\,
  \Bigg\{\left[
  \partial_{\theta}\left(\cos\theta\,\alpha_1\right)
  +\alpha_6-\frac{1}{2}\,\alpha_2\alpha_8
  +\alpha_4\alpha_6-\frac{1}{2}\,\cos^2\theta\,\alpha_3\alpha_7
  \right]^2 %%\right.
  \nonumber \\
  &+ \left[
  \partial_{\theta}\alpha_2
  -\cos\theta\,\alpha_7
  -\frac{1}{2}\cos\theta\left(\alpha_3\alpha_6
              -\alpha_1\alpha_8-\sqrt{3}\,\alpha_5\alpha_7-\alpha_4\alpha_7\right)
  \right]^2
  \nonumber \\
  &+ \left[\partial_{\theta}\left(\cos\theta\,\alpha_3\right)-  2\,\alpha_8
   -\frac{1}{2}\,\alpha_4\alpha_8
   +\frac{1}{2}\,\alpha_2\alpha_6
  +\frac{1}{2}\sqrt{3}\,\alpha_5\alpha_8
  +\frac{1}{2}\cos^2\theta\,\alpha_1\alpha_7 \right]^2
  \nonumber \\
  &+ \left[\partial_{\theta}\alpha_4
  -\cos\theta\left(\alpha_1\alpha_6
      +\frac{1}{2}\,\alpha_2\alpha_7
      -\frac{1}{2}\,\alpha_3\alpha_8\right)\right]^2 \nonumber \\
  &  %%\left.
  + \left[
  \partial_{\theta}\alpha_5
  -\frac{\sqrt{3}}{2}\cos\theta\left(\alpha_3\alpha_8
                             +\alpha_2\alpha_7\right)
   \right]^2 \Bigg\},
\end{align}

\begin{align}
\label{eq:edens-Hkin}
  \widehat{e}_\textrm{\,Hkin}&=
  \eta^2\,
  \bigg\{\left(\partial_r\beta_1\right)^2
  + \cos^2\theta\left(\partial_r\beta_2\right)^2
  + \left(\partial_r\beta_3\right)^2 \bigg\}
  \nonumber \\
  &+
  \frac{\eta^2}{r^2}\,
  \left\{\left[\partial_{\theta}\beta_1
  -\frac{1}{2}\cos\theta\left(\alpha_7\beta_3+\alpha_6\beta_2\right)\right]^2
  +\left[\partial_{\theta}\left(\cos\theta\,\beta_2\right)
  +\frac{1}{2}\left(\alpha_6\beta_1-\alpha_8\beta_3\right)\right]^2\right.
  \nonumber \\
  &\left.\qquad\quad+
  \left[\partial_{\theta}\beta_3+\frac{1}{2}\cos\theta\left(\alpha_8\beta_2
  +\alpha_7\beta_1\right)\right]^2\right\}
  \nonumber \\
  &+
  \frac{\eta^2}{4r^2\sin^2\theta}\,
  \left\{\left[\alpha_4\beta_1
  +\alpha_5\beta_1/\sqrt{3}+\cos^2\theta\,\alpha_1\beta_2
  +\alpha_2\beta_3\right]^2\right.
  \nonumber \\
  &\qquad\qquad\quad+\cos^2\theta\left[2\,\beta_2-\alpha_1\beta_1
  +\alpha_4\beta_2-\alpha_5\beta_2/\sqrt{3}-\alpha_3\beta_3\right]^2 \nonumber \\
  &\left.\qquad\qquad\quad+\left[2\,\beta_3+\alpha_2\beta_1
  -2\alpha_5\beta_3/\sqrt{3}+\cos^2\theta\,\alpha_3\beta_2\right]^2\right\}, %\\
  %& \nonumber \\
  \end{align}

\begin{align}
  \label{eq:edens-Hpot}
  \widehat{e}_\textrm{\,Hpot}=&\lambda\,
  \eta^4
  \left(\beta_1^2+\cos^2\theta\,\beta_2^2  +\beta_3^2-1\right)^2.
\end{align}

%%\newpage%%tmp
\section{Expansion coefficients for the Ansatz functions
in the basic $SU(3)$ YMH theory}
\label{app:Expansion-coefficients-basic-YMHth}

In this appendix, we give the details of the double expansion
of the $\widehat{S}$ \textit{Ansatz} functions. In view of
the behavior \eqref{alphas-betas-origin}
at the origin and the boundary conditions
\eqref{alphaBCSinfinity} and \eqref{betaBCSinfinity}
towards spatial infinity,
we redefine the two-dimensional profile functions of the generalized \textit{Ansatz} as follows:
\bsubeqs\label{eq:bar_profs_def}
\bea
  \left(\begin{array}{c}
  	\overline{\alpha}_1(x,\,\theta) \\[2mm]
  	\overline{\alpha}_2(x,\,\theta) \\[2mm]
  	\overline{\alpha}_3(x,\,\theta) \\[2mm]
  	\overline{\alpha}_4(x,\,\theta) \\[2mm]
  	\overline{\alpha}_5(x,\,\theta) \\[2mm]
  	\overline{\alpha}_6(x,\,\theta) \\[2mm]
  	\overline{\alpha}_7(x,\,\theta) \\[2mm]
  	\overline{\alpha}_8(x,\,\theta)
  \end{array}\right) &=& \left(\begin{array}{c}
    \alpha_1(x,\,\theta)/[-4x^2\sin\theta] \\[2mm]
    \alpha_2(x,\,\theta)/[2x^2\sin\theta] \\[2mm]
    \alpha_3(x,\,\theta)/[-2x^3\sin^2\theta] \\[2mm]
    \alpha_4(x,\,\theta)/[-3x^2\sin^2\theta] \\[2mm]
    \alpha_5(x,\,\theta)/[\sqrt{3}x^2\sin^2\theta] \\[2mm]
    \alpha_6(x,\,\theta)/[2x^2] \\[2mm]
    \alpha_7(x,\,\theta)/[2x^2] \\[2mm]
    \alpha_8(x,\,\theta)/[-2x^3\sin\theta]
  \end{array}\right),
\eea
\bea
	\left(\begin{array}{c}
		\overline{\beta}_1(x,\,\theta) \\[2mm]
		\overline{\beta}_2(x,\,\theta) \\[2mm]
		\overline{\beta}_3(x,\,\theta)	
	\end{array}\right) &=& \left(\begin{array}{c}
		\beta_1(x,\,\theta)/[x] \\[2mm]
		\beta_2(x,\,\theta)/[-x^2\sin\theta] \\[2mm]
		\beta_3(x,\,\theta)/[-x\sin\theta]
	\end{array}\right).
\eea\esubeqs
These redefinitions rely on seven symmetry-axis boundary conditions,
given by \eqref{alphaBCSaxis12},  \eqref{alphaBCSaxis345},
and \eqref{betaBCSaxis23}.
The four remaining boundary conditions on the symmetry axis
($\overline{\theta}=0,\pi$) are%
\bsubeqs\label{eq:sym_bcs_hat_profs}
\bea
\overline{\alpha}_6(x,\overline{\theta}) &=&
2\cos\theta\,\partial_{\theta}\big[\sin\theta\,\overline{\alpha}_1(x,\,\theta)\big]\Big|_{\theta=\overline{\theta}}\,,
  \\[2mm]
\overline{\alpha}_7(x,\overline{\theta}) &=&
\cos\theta\,\partial_{\theta}\big[\sin\theta\, \overline{\alpha}_2(x,\,\theta)\big]\Big|_{\theta=\overline{\theta}}\,,
  \\[2mm]
\overline{\alpha}_8(x,\overline{\theta}) &=&
\left(\cos^2\theta\, \overline{\alpha}_3
+\frac{1}{2}\sin\theta\cos\theta\,\partial_{\theta}\overline{\alpha}_3\right) \bigg|_{\theta=\overline{\theta}} \,,
  \\[2mm]
  \partial_{\theta} \overline{\beta}_1(x,\,\theta) \Big|_{\theta=\overline{\theta}} &=& 0 \,.
\eea\esubeqs
The boundary conditions of the redefined \textit{Ansatz} functions at
spatial infinity take values in the range $[0,1]$,
\bsubeqs\label{eq:S_hat_bar_{i}nfbcs}
\bea
\lim_{x\rightarrow 1}\left(\begin{array}{c}
\overline{\alpha}_1(x,\,\theta) \\[2mm]
\overline{\alpha}_2(x,\,\theta) \\[2mm]
\overline{\alpha}_3(x,\,\theta) \\[2mm]
  																\overline{\alpha}_4(x,\,\theta) \\[2mm]
  																\overline{\alpha}_5(x,\,\theta) \\[2mm]
  																\overline{\alpha}_6(x,\,\theta) \\[2mm]
  																\overline{\alpha}_7(x,\,\theta) \\[2mm]
  																\overline{\alpha}_8(x,\,\theta)
\end{array}\right)
&=&
\left(\begin{array}{c}
  																(1+\sin^2\theta)/2 \\[2mm]
  																\cos^2\theta \\[2mm]
  																1 \\[2mm]
  																(1+2\sin^2\theta)/3 \\[2mm]
  																1 \\[2mm]
  																1 \\[2mm]
  																1 \\[2mm]
  																1
\end{array}\right),
  \eea
  \bea
  \lim_{x\rightarrow 1}\left(\begin{array}{c}
  																\overline{\beta}_1(x,\,\theta) \\[2mm]
  																\overline{\beta}_2(x,\,\theta) \\[2mm]
  																\overline{\beta}_3(x,\,\theta)
  																\end{array}\right)&=&
  																\left(\begin{array}{c}
  																\cos^2\theta \\[2mm]
  																1 \\[2mm]
  																1
\end{array}\right).
\eea\esubeqs

We now expand these redefined \textit{Ansatz} functions,
first in $\theta$ and then in $x$.
Specifically, the $\theta$ expansion is given by
\bsubeqs\label{eq:S_hat_angular_expansion_1_2}
\bea\label{eq:S_hat_angular_expansion_1}
  \overline{\alpha}_{i}(x,\,\theta)&=&
  \frac{f_{i0}(x)}{2} + \sum_{n=1}^N \Big[f_{in}(x) \cos(2n\theta) + p_{in}(x) \sin([2n-1]\theta)\Big]
\nonumber\\
&&+ \begin{cases} p_{20}(x)\, |\cos\theta|,\qquad\ \text{for } i=2, \\[.5ex]
p_{70}(x)/|\cos\theta|,\qquad \text{for } i=7, \\[.5ex]
0,\qquad\qquad\qquad\quad\ \text{for } i=1,3,4,5,6,8,
  \end{cases}
\eea
\bea
\label{eq:S_hat_angular_expansion_2}
  \overline{\beta}_{j}(x,\,\theta)&=&
  \frac{h_{j0}(x)}{2} + \sum_{n=1}^N \Big[h_{jn}(x) \cos(2n\theta) + q_{jn}(x) \sin([2n-1]\theta)\Big]
\nonumber\\
 &&+ \begin{cases} q_{10}(x)\, |\cos\theta|,\quad \text{for } j=1, \\[.5ex]
  											0,\qquad\qquad\qquad \text{for } j=2,3.
  \end{cases}
\eea\esubeqs
With the following boundary conditions at the origin:
\bsubeqs\label{eq:rad_profile_fct_bcs_origin}
\bea
  f_{i0}(0)&=&0,\quad \text{for } i=2,7,
  \\[2mm]
  h_{10}(0)&=&0,
  \\[2mm]
  f_{in}(0)&=&p_{in}(0)=h_{jn}(0)=q_{jn}(0)=0,\quad \forall i,j \text{ and } n>0,
\eea\esubeqs
expansions \eqref{eq:S_hat_angular_expansion_1} and \eqref{eq:S_hat_angular_expansion_2} yield precisely the analytically determined behavior \eqref{alphas-betas-origin}
near the origin, provided the radial functions $f(x)$, $h(x)$, $p(x)$ and $q(x)$ contain only positive powers of $x$.
It can be seen, that consistency of the expansions \eqref{eq:S_hat_angular_expansion_1} with the symmetry axis boundary conditions \eqref{eq:sym_bcs_hat_profs} also demands that
\begin{equation}
  p_{20}(x)=p_{70}(x),
\end{equation}
which we ensure by replacing $p_{70}$ with $p_{20}$ in the angular expansion.

The boundary conditions towards $x=1$, given by
\eqref{eq:S_hat_bar_{i}nfbcs},
require the following boundary conditions of our radial functions:
\bsubeqs\label{eq:rad_prof_{i}nfbcs}
\bea
  f_{in}(1) &=& \left(\begin{array}{cccccccc}
+3/2 \;&\; 1   \;&\; 2 \;&\; +4/3  \;&\; 2 \;&\; 2 \;&\; 2 \;&\; 2 \\[2mm]
-1/4 \;&\; 1/2 \;&\; 0 \;&\; -1/3  \;&\; 0 \;&\; 0 \;&\; 0 \;&\; 0
  			\end{array}\right),\quad \text{for } n\in [0,1],
\eea
\bea
h_{jn}(1) &=& \left(\begin{array}{ccc}
  				1   \;&\; 2 \;&\; 2 \\[2mm]
  				1/2 \;&\; 0 \;&\; 0
  			\end{array}\right),\quad \text{for } n\in [0,1],
\\[2mm]
  f_{in}(1) &=& 0,\quad h_{jn}(1) = 0,\quad \text{for } n>1 \,,
\\[2mm]
p_{in}(1) &=& 0,\quad q_{jn}(1) = 0,\quad \text{for } n\geq 0.
\eea\esubeqs
In addition, we must account for the four boundary conditions \eqref{eq:sym_bcs_hat_profs}
on the symmetry axis. We do this by fixing the radial profile functions $f_{60}$, $f_{70}$, $f_{80}$ and $q_{11}$
from the following conditions:
\bsubeqs
\bea
	\frac{f_{60}-2f_{10}}{2} + \sum_{n=1}^N\left[f_{6n} - 2f_{1n}\right]
&=& 0,
\\[2mm]
	\frac{f_{i,0}-f_{i-5,0}}{2} + \sum_{n=1}^N\left[f_{i,n} - f_{i-5,n}\right]
&=& 0,\quad \text{for } i=7,8,
\\[2mm]
	\sum_{n=1}^N(2n-1)\,q_{1n}
&=& 0.
\eea\esubeqs

We next expand the obtained radial functions in Legendre polynomials
$P_m(2x-1)$ [these polynomials are normalized to $P_m(1)=1$ and
orthogonal over $x\in [0,1]$  with weight $1$]:
\bsubeqs\label{eq:S_hat_radial_expansion}
\bea
  f_{in}(x) &=& x^2\sum_{m=0}^M a_{inm}\,P_m(2x-1)
+ \left\{\begin{array}{ll}
  2e_{i},     &\text{for\;\;} i=1,3,4,5 \text{ and } n=0,
  \\
  2e_{i-5}, &\text{for\;\;} i=6,8 \text{ and } n=0,
  \\
  0,        &\text{for\;\;} i=2,7 \text{ or } n>0,
  \end{array}\right.
\\[2mm]
  h_{jn}(x) &=& x^2\sum_{m=0}^M b_{jnm}\,P_m(2x-1)
+ \left\{\begin{array}{ll}
  2e_{j+5}, &\text{for\;\;} j=2,3, \text{ and } n=0,
  \\
  0,        &\text{for\;\;} j=1 \text{ or } n>0,
  \end{array}\right.
\eea
\bea											
  p_{in}(x) &=& x^2\sum_{m=0}^M c_{inm}\,P_m(2x-1)
+ \left\{\begin{array}{ll}
  e_2,  &\quad\text{for\;\;} i=2 \text{ and } n=0,
  \\
  0,    &\quad\text{for\;\;} n>0,
  \end{array}\right.
\\[2mm]	 												
q_{jn}(x) &=& x^2\sum_{m=0}^M d_{jnm}\,P_m(2x-1)
+ \left\{\begin{array}{ll}
  e_{6}, &\quad\text{for\;\;} j=1, \text{ and } n=0,
  \\
  0,     &\quad\text{for\;\;} n>0,
  \end{array}\right.
\eea\esubeqs
where the eight coefficients $e_{k}$ are proportional
to the eight origin coefficients from \eqref{alphas-betas-origin}.
The $x^2$ prefactors in \eqref{eq:S_hat_radial_expansion}
ensure that the boundary conditions \eqref{eq:rad_profile_fct_bcs_origin}
at the origin are always met, regardless of the values the expansion coefficients
may take
during the minimization process. Only the boundary conditions \eqref{eq:rad_prof_{i}nfbcs} at $x=1$ require fixing during minimization.
This is easily done by adjusting one expansion coefficient of each radial function expansion in the following conditions:
\bsubeqs %\label{eq:}
\bea
\sum_{m=0}^M a_{inm} &=&
f_{in}(1)
- \left\{\begin{array}{ll}
  2\,e_{i},    &\text{for\;\;} i=1,3,4,5 \text{ and } n=0,
  \\
  2\,e_{i-5},&\text{for\;\;} i=6,8 \text{ and } n=0,
  \\
  0,       &\text{for\;\;} i=2,7 \text{ or } n>0,
  \end{array}\right.
\\[2mm]
\sum_{m=0}^M b_{jnm} &=&
 h_{jn}(1)
- \left\{\begin{array}{ll}
  2\,e_{j+5}, &\text{for\;\;} j=2,3, \text{ and } n=0,
  \\
  0,        &\text{for\;\;} j=1 \text{ or } n>0,
  \end{array}\right.
\\[2mm]  												
\sum_{m=0}^M c_{inm} &=&
p_{in}(1)
- \begin{cases}
  										  e_2, \qquad \text{for } i=2 \text{ and } n=0, \\
  										  0, \qquad\ \text{for } n>0,
\end{cases}
\\[2mm]  	 												  											  \sum_{m=0}^M d_{jnm} &=&
q_{jn}(1)
- \begin{cases}
  										  e_{6}, \qquad \text{for } j=1, \text{ and } n=0, \\
  										  0, \qquad\ \text{for } n>0.
\end{cases}
\eea\esubeqs

Cutting off both expansions
at given $N$ (for $\theta$) and $M$ (for $x$), we obtain a finite set of expansion coefficients over which we can minimize. Specifically, we minimize over all $a_{imn}$ and $b_{jmn}$ in the range $n\in [0,N]$, with the exception of $a_{60m}$, $a_{70m}$ and $a_{80m}$, which are fixed by the symmetry axis conditions for all $m$.
In addition, we minimize over all $c_{imn}$ and $d_{jmn}$
in the ranges $n\in [1,N]$ and $m\in [1,M]$,
with the exception of $d_{11m}$,
while the $m=0$ coefficients are fixed by the boundary conditions at $x=1$. Finally, we also minimize over the eight origin coefficients $e_{k}$ and the coefficients $c_{20m}$ and $d_{10m}$.
This, then, gives the following total number of coefficients:
\begin{equation}\label{eq:N-coeff-basic-SU3-YMHth}
  N_{\text{coeff}}^{(\text{basic\;YMHth})}
  = 8 + \Big[11\,(2\,N+1)-2\Big]M \,,
\end{equation}
which asymptotically goes as $22\,N\,M$ for $N,\,M\to \infty$.

%%\newpage%%tmp
\section{Noncontractible sphere of configurations
in the extended $SU(3)$ YMH theory}
\label{app:NCS-in-extended-SU3-YMHth}

The basic idea behind the $\widehat{S}$ construction
has been sketched in Sec.~\ref{subsec:Minimax-procedure}.
The relevant noncontractible sphere~(NCS) of configurations
is based on the $SU(3)$ matrix
$U(\psi,\, \mu,\, \alpha,\,\theta,\, \phi)$
as given by Eqs.~(3.1) and (3.2) of Ref.~\cite{KlinkhamerRupp2005},
where the coordinates
$(\psi,\, \mu,\, \alpha)$ parameterize the 3-sphere in configuration space
and the coordinates $(\theta,\, \phi)$ refer to 2-sphere at
spatial infinity.
The matrix at the ``bottom'' of the NCS ($\psi=0$) is given by
\begin{equation}
\label{eq:V}
\hspace*{-5mm}
U(0,\, \mu,\, \alpha,\,\theta,\, \phi) =
\left(\begin{array}{ccc}
1& 0 & 0  \\
0& 0 & -1  \\
0& +1& 0
  \end{array}\right)
\equiv V  \,,
\end{equation}
whereas the matrix at the ``top''  of the NCS ($\psi=\pi$) is given by
\begin{eqnarray}
\label{eq:W}
\hspace*{-5mm}
U(\pi,\, \mu,\, \alpha,\,\theta,\, \phi) &=&
%\nonumber\\[2mm]&=&
\left(\begin{array}{ccc}
\cos^2\theta
&\;\; -\cos\theta\, \sin\theta\, e^{ i \phi}
&\;\;\sin\theta \, e^{- i \phi}
\\
-\cos\theta\, \sin\theta\, e^{ i \phi}
&\;\; \sin^2\theta \, e^{2 i \phi}
&\;\; \cos\theta
\\
-\sin\theta\, e^{- i \phi}
&\;\; -\cos\theta
&\;\; 0
  \end{array}\right)
\equiv W(\theta,\,\phi)\,.
\end{eqnarray}

%%%%%%\newpage%%tmp
The field configurations of the NCS in the extended $SU(3)$ YMH theory
have the same gauge fields as in Ref.~\cite{KlinkhamerRupp2005},
\begin{subequations}\label{eq:NCS-in-extended-SU3-YMHth}
\begin{eqnarray}
\label{eq:NCS-in-extended-SU3-YMHth-A0}
g\,A_0(r,\theta,\phi)^\text{(NCS)}_{(\zeta)} &=& 0 \,,
\\[2mm]
\label{eq:NCS-in-extended-SU3-YMHth-Am}
g\,A_m(r,\theta,\phi)^\text{(NCS)}_{(\zeta)} &=&
-f(r) \,\partial_m U(\zeta,\, \theta,\phi)\: U^{-1}(\zeta,\, \theta,\phi)\,,
\end{eqnarray}
and the following set of Higgs fields:
\begin{eqnarray}
\Phi_{1}(r,\theta,\phi)^\text{(NCS)}_{(\zeta)}
&=& h_{1}(r)\,U(\zeta,\, \theta,\,\phi)\,
\left(\begin{array}{c}
   \eta\\0\\0
\end{array}\right)
+ \left[1-h_{1}(r)\right]\,\cos^2\frac{\psi}{2}\,
\left(\begin{array}{c}
   \eta\\0\\0
\end{array}\right)\,,
\eea
\bea
\Phi_{2}(r,\theta,\phi)^\text{(NCS)}_{(\zeta)}
&=&
 h_{2}(r)\,  U(\zeta,\, \theta,\,\phi)\,M_2\,
\left(\begin{array}{c}
   \eta\\0\\0
\end{array}\right)
+ \left[1-h_{2}(r)\right]\,\cos^2\frac{\psi}{2}\,V\,M_2
\left(\begin{array}{c}
   \eta\\0\\0
\end{array}\right)
\nonumber\\
&=&
 h_{2}(r)\,  U(\zeta,\, \theta,\,\phi)\,
\left(\begin{array}{c}
   0\\0\\-\eta
\end{array}\right)
+ \left[1-h_{2}(r)\right]\,\cos^2\frac{\psi}{2}\,
\left(\begin{array}{c}
   0\\\eta\\0
\end{array}\right)\,,
\eea
\bea
\Phi_{3}(r,\theta,\phi)^\text{(NCS)}_{(\zeta)}
&=&
 h_{3}(r)\,  U(\zeta,\, \theta,\,\phi)\,M_3^\dagger\,
\left(\begin{array}{c}
   \eta\\0\\0
\end{array}\right)
+ \left[1-h_{3}(r)\right]\,\cos^2\frac{\psi}{2}\,V\,M_3^\dagger
\left(\begin{array}{c}
   \eta\\0\\0
\end{array}\right)
\nonumber\\
&=&
 h_{3}(r)\,  U(\zeta,\, \theta,\,\phi)\,
\left(\begin{array}{c}
   0\\\eta\\0
\end{array}\right)
+ \left[1-h_{3}(r)\right]\,\cos^2\frac{\psi}{2}\,
\left(\begin{array}{c}
   0\\0\\ \eta
\end{array}\right)\,,
\end{eqnarray}
\end{subequations}
with the short-hand notation $\zeta \equiv (\psi,\, \mu,\, \alpha)$
and the $SU(3)$ matrices $M_2$ and $M_3$ defined by \eqref{eq:M1-M2-M3}.
The radial functions $f(r)$ and $h_\alpha(r)$
of the NCS \eqref{eq:NCS-in-extended-SU3-YMHth}
have boundary conditions
\begin{subequations}\label{eq:NCS-in-extended-SU3-YMHth-f-h-bcs}
\begin{eqnarray}
f(0)&=&h_1(0)=h_2(0)=h_3(0)=0\,,
\\[2mm]
f(\infty)&=&h_1(\infty)=h_2(\infty)=h_3(\infty)=1\,.
\end{eqnarray}
\end{subequations}

%%%%%%%\newpage%%tmp
The NCS fields \eqref{eq:NCS-in-extended-SU3-YMHth} at $\psi=0$,
with $U=V$ from \eqref{eq:V}, are given by
\begin{subequations}\label{eq:NCS-in-extended-SU3-YMHth-vac}
\begin{eqnarray}
g\,A_0(r,\theta,\phi)^\text{(NCS)}\,\Big|_{\psi=0} &=&
0\,,
\\[2mm]
g\,A_m(r,\theta,\phi)^\text{(NCS)}\,\Big|_{\psi=0} &=&
0\,,
\eea
\bea
\Phi_{1}(r,\theta,\phi)^\text{(NCS)}\,\Big|_{\psi=0}
&=&
\left(\begin{array}{c}
   \eta\\0\\0
\end{array}\right)\,,
\\[2mm]
\Phi_{2}(r,\theta,\phi)^\text{(NCS)}\,\Big|_{\psi=0}
&=&
\left(\begin{array}{c}
   0\\\eta\\0
\end{array}\right)\,,
\\[2mm]
\Phi_{3}(r,\theta,\phi)^\text{(NCS)}\,\Big|_{\psi=0}
&=&
\left(\begin{array}{c}
   0\\0\\\eta
\end{array}\right)\,,
\end{eqnarray}
\end{subequations}
which correspond to the fields \eqref{eq:actionYMH2-Higgs-vac}
of the classical vacuum.

For nontrivial radial functions $f(r)$ and $h_\alpha(r)$
with boundary conditions
\eqref{eq:NCS-in-extended-SU3-YMHth-f-h-bcs},
the NCS fields \eqref{eq:NCS-in-extended-SU3-YMHth} at $\psi=\pi$
correspond to a first approximation of the $\widehat{S}$
fields in the extended theory.
Specifically, these fields are given by
\begin{subequations}\label{eq:NCS-in-extended-SU3-YMHth-approx-Shat}
\begin{eqnarray}
\hspace*{-10mm}
g\,A_0(r,\theta,\phi)^\text{(NCS)}\,\Big|_{\psi=\pi} &=&
0\,,
\\[2mm]
\hspace*{-10mm}
g\,A_m(r,\theta,\phi)^\text{(NCS)}\,\Big|_{\psi=\pi} &=&
-f(r) \,\partial_m W(\theta,\phi)\: W^{-1}(\theta,\phi)\,,
\eea
\bea
\label{eq:NCS-in-extended-SU3-YMHth-approx-Shat-Phi1}
\hspace*{-10mm}
\Phi_{1}(r,\theta,\phi)^\text{(NCS)}\,\Big|_{\psi=\pi}
&=&
h_{1}(r)\,W(\theta,\,\phi)\,
\left(\begin{array}{c}
   \eta\\0\\0
\end{array}\right)
=
h_{1}(r)\,\eta\,
\left(\begin{array}{c}
  \cos^2\theta\\
  -\cos\theta\,\sin\theta\;e^{ i \phi}\\
  -\sin\theta\;e^{- i \phi}
\end{array}\right)
\,,
\\[2mm]
\label{eq:NCS-in-extended-SU3-YMHth-approx-Shat-Phi2}
\hspace*{-10mm}
\Phi_{2}(r,\theta,\phi)^\text{(NCS)}\,\Big|_{\psi=\pi}
&=&
  h_{2}(r)\,W(\theta,\,\phi)\,
\left(\begin{array}{c}
   0\\0\\-\eta
\end{array}\right)
=
h_{2}(r)\,\eta\,
\left(\begin{array}{c}
   -\sin\theta\;e^{- i \phi}\\
  -\cos\theta\\
  0
\end{array}\right)
\,,
\\[2mm]
\hspace*{-10mm}
\label{eq:NCS-in-extended-SU3-YMHth-approx-Shat-Phi3}
\Phi_{3}(r,\theta,\phi)^\text{(NCS)}\,\Big|_{\psi=\pi}
&=&
 h_{3}(r)\,  W(\theta,\,\phi)\,
\left(\begin{array}{c}
   0\\ \eta \\0
\end{array}\right)
=
 h_{3}(r)\,\eta\,
\left(\begin{array}{c}
  -\cos\theta\,\sin\theta\; e^{ i \phi}\\
  \sin^2\theta\; e^{2 i \phi}\\
  -\cos\theta
\end{array}\right)
\,,
\end{eqnarray}
\end{subequations}
in terms of the $SU(3)$ matrix $W$ defined by \eqref{eq:W}.
As discussed in Sec.~\ref{subsec:Minimax-procedure},
the $\widehat{S}$ \textit{Ansatz} is obtained by a generalization
of the fields \eqref{eq:NCS-in-extended-SU3-YMHth-approx-Shat}
and is presented in Sec.~\ref{subsec:Shat-Ansatz-in-extended-SU3-YMHth}.

%%\newpage%%tmp
\section{$\widehat{S}$ energy density in the extended $SU(3)$ YMH theory}
\label{app:Shat-energy-density-extended-SU3-YMHth}

The $\widehat{S}$ \textit{Ansatz} in the
extended $SU(3)$ Yang--Mills--Higgs theory \eqref{eq:actionYMH2}
has been presented in Sec.~\ref{subsec:Shat-Ansatz-in-extended-SU3-YMHth}.
The corresponding energy density is as follows:
\begin{subequations}\label{eq:edens-in-extended-SU3-YMHth}
\begin{eqnarray}
\widehat{e}(r,\,\theta)^\text{(ext.\;YMHth)} &=&
\widehat{e}_{\rm \,YM}(r,\,\theta)
+ \widehat{e}_{\rm \,Hkin,\,123}(r,\,\theta)
+ \widehat{e}_{\rm \,Hpot,\,123}(r,\,\theta)\,,
\\[2mm]
\widehat{e}_{\rm \,Hkin,\,123}(r,\,\theta)
&=&
\widehat{e}_{\rm \,Hkin,\,1}(r,\,\theta)+
\widehat{e}_{\rm \,Hkin,\,2}(r,\,\theta)+
\widehat{e}_{\rm \,Hkin,\,3}(r,\,\theta)\,,
\end{eqnarray}
\end{subequations}
where $\widehat{e}_{\rm \,YM}$ equals the previous
result \eqref{eq:edens-YM}
and $\widehat{e}_{\rm \,Hkin,\,1}$ is identical to \eqref{eq:edens-Hkin}.
The Higgs fields $\Phi_2$ and $\Phi_3$ give the following
further contributions:
\begin{subequations}\label{eq:edens23-in-extended-SU3-YMHth}
\begin{eqnarray}
\label{eq:edens23-in-extended-SU3-YMHth-Hkin2}
\widehat{e}_{\rm \,Hkin,\,2}(r,\,\theta)&=&
\eta^2
\Bigg\{
\big[\partial_r\beta_4\big]^2
+ \cos^2\theta\, \big[\partial_r\beta_5\big]^2
+ \big[\partial_r\beta_6\big]^2
\Bigg\}
\nonumber \\
&&
+\frac{\eta^2}{4r^2}
\Bigg\{
 \Big[2\,\partial_{\theta}\beta_4-\cos\theta\, \alpha_6\beta_5\Big]^2
+\Big[2\cos\theta\, \partial_{\theta}\beta_5
       +\alpha_6\beta_4-2\sin\theta\, \beta_5\Big]^2
\nonumber \\
&&\hspace{3em}
+ 4\Big[\partial_{\theta}\beta_6\Big]^2
+ \cos^2\theta\, \Big[\alpha_7\beta_4+\alpha_8\beta_5\Big]^2
+ \Big[\cos\theta\, \alpha_7\beta_6\Big]^2
+ \Big[\alpha_8\beta_6\Big]^2
\Bigg\}
\nonumber \\
&&
+\frac{\eta^2}{12r^2\sin^2\theta}
\Bigg\{
 \left[\sqrt{3}\,\cos^2\theta\, \alpha_1\beta_5 + 2\sqrt{3}\,\beta_4
       +\sqrt{3}\,\alpha_4\beta_4+\alpha_5\beta_4\right]^2
+3\Big[\alpha_2\beta_6\Big]^2
\nonumber \\
&&\hspace{6em}
+\cos^2\theta\, \left[\sqrt{3}\,\left(\alpha_1\beta_4-\alpha_4\beta_5\right)
                   +\alpha_5\beta_5\right]^2
+ 3\cos^2\theta\, \Big[\alpha_3\beta_6\Big]^2
\nonumber \\
&&\hspace{6em}
+4\Big[\alpha_5\beta_6\Big]^2 + 3\Big[\alpha_2\beta_4
        +\cos^2\theta\, \alpha_3\beta_5\Big]^2
\Bigg\}\,,
%\\[2mm]
\end{eqnarray}
\begin{eqnarray}
\label{eq:edens23-in-extended-SU3-YMHth-Hkin3}
\widehat{e}_{\rm \,Hkin,\,3}(r,\,\theta)&=&
 \eta^2
\Bigg\{
 \cos^2\theta\, \big[\partial_r\beta_7\big]^2
 + \big[\partial_r\beta_8\big]^2
 + \cos^2\theta\, \big[\partial_r\beta_9\big]^2
\Bigg\}
\nonumber \\
&&
+\frac{\eta^2}{4r^2}
\Bigg\{
\Big[2\,\partial_{\theta}\left(\cos\theta\, \beta_7\right)
 -\alpha_6\beta_8 -\cos^2\theta\, \alpha_7\beta_9\Big]^2
\nonumber \\
&&\hspace{3em}
+\Big[2\,\partial_{\theta}\beta_8
+\cos\theta\, \alpha_6\beta_7-\cos\theta\, \alpha_8\beta_9\Big]^2
\nonumber \\
&&\hspace{3em}
+\Big[2\,\partial_{\theta}\left(\cos\theta\, \beta_9\right)
      +\alpha_8\beta_8+\cos^2\theta\, \alpha_7\beta_7\Big]^2
\Bigg\}
\nonumber \\
&&
+\frac{\eta^2}{12r^2\sin^2\theta}
\Bigg\{
\cos^2\theta\, \left[\sqrt{3}\left(\alpha_1\beta_8 +\alpha_2\beta_9\right)
+\frac{\beta_7}{\sqrt{3}}\left(3\,\alpha_4+\sqrt{3}\,\alpha_5
                   -6\right)\right]^2
\nonumber \\
&&\hspace{6em}
+\left[\cos^2\theta\, \sqrt{3}\left(\alpha_1\beta_7+\alpha_3\beta_9\right)
- \frac{\beta_8}{\sqrt{3}}\left(3\,\alpha_4-\sqrt{3}\,\alpha_5+12\right)\right]^2 
\nonumber \\
&&\hspace{6em}
+\cos^2\theta\, \left[\sqrt{3}\,\alpha_2\beta_7+\sqrt{3}\,\alpha_3\beta_8
                   -2\,\alpha_5\beta_9\right]^2
\Bigg\}\,.
%\\[2mm]
\end{eqnarray}
\end{subequations}
The potential energy density from the three Higgs triplets
is given by
\begin{eqnarray}
\label{eq:edens23-in-extended-SU3-YMHth-Hpot123}
\widehat{e}_{\rm \,Hpot,\,123}(r,\,\theta)&=&
\lambda\, \eta ^4  \Bigg\{
\Big[\beta_1^2+\cos ^2\theta\, \beta_2^2+\beta_3^2-1\Big]^2
+\Big[\beta_4^2+\cos ^2\theta\, \beta_5^2+\beta_6^2-1\Big]^2
\nonumber \\
&&\hspace{3em}
+ \Big[\cos ^2\theta\, \beta_7^2 + \beta_8^2
+ \cos^2\theta\, \beta_9^2-1\Big]^2
+\Big[\beta_1 \beta_4+\cos ^2\theta\, \beta_2 \beta_5\Big]^2
\nonumber \\
&&\hspace{3em}
+\big[\beta_3 \beta_6\big]^2
+ \cos ^2\theta\Big[\beta_1 \beta_7+\beta_2 \beta_8+\beta_3 \beta_9\Big]^2
\nonumber \\
&&\hspace{3em}
+ \cos ^2\theta\Big[\beta_4 \beta_7+\beta_5 \beta_8\Big]^2
+ \cos ^2\theta\Big[\beta_6 \beta_9\Big]^2
\Bigg\}\,.
\end{eqnarray}

%%\newpage%%tmp
\section{Minimization setup in the extended $SU(3)$ YMH theory}
\label{app:Minimization-setup-extended-SU3-YMHth}

The numerical  minimization procedure for $\widehat{S}$
in the extended $SU(3)$ YMH theory \eqref{eq:actionYMH2}
is similar to the one
in the basic $SU(3)$ YMH theory \eqref{eq:actionYMH1}.
The procedure for the Yang-Mills
\textit{Ansatz} functions $\alpha_{i}$ ($i=1,\,\ldots,\,8$)
and the Higgs \textit{Ansatz} functions $\beta_{k}$ ($k=1,\,2,\,3$)
remains unchanged.
Their expansion coefficients and constraints are given in
Appendix~\ref{app:Expansion-coefficients-basic-YMHth}.

In view of the behavior \eqref{eq:beta456789-origin}
at the origin and the boundary conditions
\eqref{eq:Shat-in-extended-SU3-YMHth-Ansatz-Higgs-fields-bcs-infty}
towards spatial infinity,
we redefine, by analogy with \eqref{eq:bar_profs_def},
the further profile functions:
\begin{equation}\label{eq:bar_profs_def_ext}
	\left(\begin{array}{c}
		\overline{\beta}_4(x,\,\theta) \\[2mm]
		\overline{\beta}_5(x,\,\theta) \\[2mm]
		\overline{\beta}_6(x,\,\theta) \\[2mm]
		\overline{\beta}_7(x,\,\theta) \\[2mm]
		\overline{\beta}_8(x,\,\theta) \\[2mm]
		\overline{\beta}_9(x,\,\theta)	
	\end{array}\right) = \left(\begin{array}{c}
		\beta_4(x,\,\theta)/[-x\sin\theta] \\[2mm]
		\beta_5(x,\,\theta)/[-x] \\[2mm]
		\beta_6(x,\,\theta)/[x] \\[2mm]
		\beta_7(x,\,\theta)/[-x^2\sin\theta] \\[2mm]
		\beta_8(x,\,\theta)/[x^2\sin^2\theta] \\[2mm]
		\beta_9(x,\,\theta)/[-x]
	\end{array}\right).
\end{equation}
These redefinitions rely on three symmetry-axis boundary conditions, given
by \eqref{eq:ext-YMHth-Ansatz-Higgs-bcs-axis2347}
for $k=4,\,7$,
and \eqref{eq:ext-YMHth-Ansatz-Higgs-bcs-axis8}
for $k=8$.
The three remaining boundary conditions on the symmetry axis
($\overline{\theta}=0,\pi$) are
\begin{equation}\label{eq:sym_bcs_hat_profs_ext}
  \partial_{\theta} \overline{\beta}_{k}(x,\,\theta) \Big|_{\theta=\overline{\theta}} = 0\quad \text{for }k=5,6,9.
\end{equation}
The boundary conditions of the redefined \textit{Ansatz} functions at
spatial infinity are then
%\bsubeqs\label{eq:}
%\bea\eea\esubeqs
\begin{equation}\label{eq:S_hat_bar_{i}nfbcs_ext}
  \lim_{x\rightarrow 1}\left(\begin{array}{c}
\overline{\beta}_4(x,\,\theta) \\[2mm]
\overline{\beta}_5(x,\,\theta) \\[2mm]
\overline{\beta}_6(x,\,\theta) \\[2mm]
\overline{\beta}_7(x,\,\theta) \\[2mm]
\overline{\beta}_8(x,\,\theta) \\[2mm]
\overline{\beta}_9(x,\,\theta)
\end{array}\right)=
\left(\begin{array}{c}
1 \\[2mm]
1 \\[2mm]
0 \\[2mm]
1 \\[2mm]
1 \\[2mm]
1
\end{array}\right).
\end{equation}
Almost identical to \eqref{eq:S_hat_angular_expansion_2},
we define the following
angular expansions of the redefined \textit{Ansatz} functions:
\bea\label{eq:S_hat_angular_expansion_ext}
  \overline{\beta}_{k}(x,\,\theta)
  &=&
  \frac{h_{k0}(x)}{2} + \sum_{n=1}^N \Big[h_{kn}(x) \cos(2n\theta) + q_{kn}(x) \sin([2n-1]\theta)\Big]
\nonumber\\[1mm]
&&
  + \begin{cases} q_{60}(x)\, |\cos\theta|,\quad \text{for } k=6, \\[.5ex]
  											0,\qquad\qquad\qquad \text{for } k=4,5,7,8,9, \end{cases}
\eea
with the following boundary conditions at the origin:
\bsubeqs\label{eq:rad_profile_fct_bcs_origin_ext}
\bea
  h_{60}(0)&=&0,
\\[2mm]
  h_{kn}(0)&=&q_{kn}(0)=0,\quad \forall i,k \text{ and } n>0.
\eea\esubeqs
With these constraints, the profile functions behave as \eqref{eq:beta456789-origin}  near the origin.
The boundary conditions at $x=1$ translate to those of the radial functions $h(x)$ and $q(x)$,
\bsubeqs\label{eq:rad_prof_{i}nfbcs_ext}
\bea
h_{kn}(1)
  &=& \left(\begin{array}{cccccc}
  				2 \;&\; 2 \;&\; 0 \;&\; 2 \;&\; 2 \;&\; 2 \\[2mm]
  				0 \;&\; 0 \;&\; 0 \;&\; 0 \;&\; 0 \;&\; 0
  			\end{array}\right),\quad
  \text{for }  k=4,\dots,9 \text{ and } n\in [0,1]\,,
\\[2mm]
h_{kn}(1) &=& 0,\quad \text{for } n>1 \,,
\\[2mm]
q_{kn}(1) &=& 0,\quad \text{for } n\geq 0\,.
\eea\esubeqs
The three boundary conditions \eqref{eq:sym_bcs_hat_profs_ext} on the symmetry axis are implemented
by fixing the radial profile functions $q_{51}$, $q_{61}$ and $q_{91}$
from the following equations:
\begin{equation}
	\sum_{n=1}^N(2n-1)\,q_{kn} = 0\,,
\quad \text{for } k=5,6,9\,.
\end{equation}

We next expand the radial functions in Legendre polynomials $P_m(2x-1)$,
\bsubeqs\label{eq:S_hat_radial_expansion_ext}
\bea
h_{kn}(x) &=& x^2\sum_{m=0}^M b_{knm}\,P_m(2x-1)
+
  \left\{\begin{array}{ll}
  2\,e_{k+5},  &\text{for\;\;} k=4,5,7,8,9 \text{ and } n=0,
  \\
  0,         & \text{for\;\;} k=6 \text{ or } n>0,
 \end{array}\right.
\\[2mm]   												  												  											
q_{kn}(x) &=& x^2\sum_{m=0}^M d_{knm}\,P_m(2x-1)
+
  \left\{
  \begin{array}{ll}
  e_{11},  &\quad\text{for\;\;} k=6 \text{ and } n=0,
  \\
  0,       &\quad\text{for\;\;} n>0,
  \end{array}\right.
\eea\esubeqs
with $x^2$ prefactors to ensure that the correct origin behavior is reproduced.
We enforce the boundary conditions \eqref{eq:rad_prof_{i}nfbcs_ext} at $x=1$ by adjusting one expansion coefficient of each radial function expansion
in the following conditions:
\bsubeqs%\label{eq:}
\bea
\sum_{m=0}^M b_{knm} &=&
h_{kn}(1)
- \left\{\begin{array}{ll}
  	2\,e_{k+5}, &\quad\text{for\;\;} k=4,5,7,8,9 \text{\;\;and\;\;} n=0,
  \\
  	0,          &\quad\text{for\;\;} k=6 \text{\;\;or\;\;} n>0,
  \end{array}\right.
\\[2mm]
\sum_{m=0}^M d_{knm} &=&
q_{kn}(1)
- \left\{\begin{array}{ll}
e_{11}, &\quad\quad \text{for\;\;} k=6 \text{\;\;and\;\;} n=0,
\\
0,      &\quad\quad \text{for\;\;} n>0.
\end{array}\right.
\eea\esubeqs
The total number of coefficients
for given radial ($M$) and angular ($N$) expansion cut-offs
is given by
\begin{equation}
\label{eq:N-coeff-extended-SU3-YMHth}
  N_{\text{coeff}}^{(\text{ext.\;YMHth})}
  = 14 + \Big[17\,(2\,N+1)-5\Big]\,M \,,
 \end{equation}
which asymptotically goes as $34\,N\,M$ for $N,\,M\to \infty$.

\end{appendix}

\end{document}